\documentclass{JHEP3}
\usepackage[english]{babel}
\usepackage{amssymb}
\usepackage{amsfonts}
\usepackage{amsmath}
\usepackage{graphicx}
\usepackage{epsfig}
\usepackage{cite}
\usepackage{psfrag}
\newcommand{\cA}{{\cal A}}  
  
  \newcommand{\cF}{{\cal F}}

  \newcommand{\cL}{{\cal L}}

\newcommand{\be}{\begin{equation}} \newcommand{\ee}{\end{equation}}
\newcommand{\bea}{\begin{eqnarray}} \newcommand{\eea}{\end{eqnarray}}
\newcommand{\beann}{\begin{eqnarray*}}  \newcommand{\eeann}{\end{eqnarray*}}
\newcommand{\bfig}{\begin{figure}} \newcommand{\efig}{\end{figure}}
\newcommand{\ba}{\begin{array}} \newcommand{\ea}{\end{array}}
\newcommand{\bcen}{\begin{center}} \newcommand{\ecen}{\end{center}}
\newcommand{\btab}{\begin{tabular}} \newcommand{\etab}{\end{tabular}}

\def\O{\operatorname{O}}

\newtheorem{Proposition}{Proposition}[section]

\newtheorem{Theorem}{Theorem}[section]
\newtheorem{Lemma}{Lemma}[section]
\newtheorem{Corrolary}{Corrolary}[section]

\newcommand{\bp}{\begin{Proposition}}	\newcommand{\ep}{\end{Proposition}}
\newcommand{\bt}{\begin{Theorem}}	\newcommand{\et}{\end{Theorem}}
\newcommand{\bl}{\begin{Lemma}}		\newcommand{\el}{\end{Lemma}}
\newcommand{\bc}{\begin{Corrolary}}	\newcommand{\ec}{\end{Corrolary}}
\title{Hydrodynamics of Holographic Superconductors}
\author{Irene Amado, Matthias Kaminski and Karl Landsteiner\\
   Instituto de F\'{\i}sica Te\'orica CSIC/UAM\\
   C-XVI Universidad Aut\'onoma de Madrid\\
  E-28049 Madrid, Spain\\
  E-mail: \email{Irene.Amado, Matthias.Kaminski, Karl.Landsteiner@uam.es}\\
}
\abstract{We study the poles of the retarded Green functions of 
a holographic superconductor. The model shows a second order phase
transition where a charged scalar operator condenses and a $U(1)$ symmetry is
spontaneously broken. The poles of the holographic Green functions are the quasinormal modes in an AdS black hole background. We study the spectrum of quasinormal frequencies in the broken phase, where we establish
the appearance of a massless or hydrodynamic mode at the critical temperature as expected for a second order phase transition. In the broken phase we find the pole representing second sound.
We compute the speed of second sound and its attenuation length  as function of the temperature. 
In addition we find a pseudo diffusion mode, whose frequencies are purely imaginary but with a non-zero gap at 
zero momentum. This gap goes to zero at the critical temperature.
As a technical side result we explain how to calculate holographic Green functions and their quasinormal modes for a set of operators that mix under the RG flow.}

\preprint{IFT-UAM/CSIC-09-14} 
\keywords{Holography, Superconductivity}

\begin{document}
%
%%%%%%%%%%%%%%%%%%%%%%%%%%%%%%%%%%%%%%
%%%%%%%%%%%% INTRODUCTION %%%%%%%%%%%%
%%%%%%%%%%%%%%%%%%%%%%%%%%%%%%%%%%%%%%
\section{\label{sec:intro}Introduction}
The AdS/CFT correspondence is a tool for studying  field theories in the strong coupling regime \cite{Maldacena:1997re,Gubser:1998bc,Witten:1998zw}.
The range of physical phenomena to which it can be applied is constantly increasing. One of the latest additions is the realization of spontaneous symmetry breaking and the appearance of a superfluid (often described as superconducting) phase at low temperature. A model with a charged scalar condensing in the background of a charged AdS black hole has first been introduced in \cite{Gubser:2008px}. Shortly afterwards it was realized that a charged scalar condenses as well when the black hole is neutral and it was shown explicitly that the DC conductivity is infinite in the broken phase \cite{Hartnoll:2008vx}.

By now there is a large variety of holographic models of superfluidity/superconductivity \cite{Albash:2008eh, Nakano:2008xc, Ammon:2008fc, Wen:2008pb, Roberts:2008ns,Maeda:2008ir, Basu:2008st, Basu:2008bh, Horowitz:2008bn, Gubser:2008pf, O'Bannon:2008bz, Rebhan:2008ur,Evans:2008nf, Denef:2009tp,Koutsoumbas:2009pa,Ammon:2009fe}.
Specifically hydrodynamical behavior in these models has been addressed before in \cite{Herzog:2008he,Yarom:2009uq}
where the speed of sound has been calculated from derivatives of thermodynamic quantities.  The hydrodynamic poles
of retarded Green functions have been studied in an analytical approximation for infinitesimal condensate in a
$p$-wave model in \cite{Herzog:2009ci}

In this work we are interested in the hydrodynamics of the holographic superconductor introduced in\cite{Hartnoll:2008vx}. In general, hydrodynamic behavior is connected either to the presence of a conserved charge, a spontaneously broken symmetry or a second order phase transition. As we will see in our model all three possibilities are realized. In holographic models the hydrodynamic modes appear as quasinormal modes in the AdS black hole background whose frequency vanishes in the zero momentum limit \cite{Policastro:2002se,Policastro:2002tn}. Quasinormal modes play an important role in the physics of black holes and a very good review of their properties in asymptotically flat black holes is \cite{Nollert:1999ji}. In the asymptotically flat situation the quasinormal modes are defined as solutions of linearized wave equations with purely infalling boundary conditions on the horizon outgoing ones at the boundary. In asymptotically AdS spaces the situation is somewhat different. On the horizon one still imposes infalling boundary conditions but on the conformal boundary of anti de Sitter space several possibilities of choosing boundary conditions arise: Dirichlet or Neumann or Robin (mixed) ones. The holographic interpretation fixes this degeneracy of boundary conditions by defining the quasinormal modes as the poles of the holographic Green functions \cite{Birmingham:2001pj,Starinets:2002br,Nunez:2003eq}. This of course implies to know the holographic Green functions, which are computed using the prescription given in \cite{Son:2002sd}. In situations in which there are several fields whose linearized wave equations form a coupled system of differential equations possibly subject to a constraint due to a gauge symmetry the construction of the holographic Green functions is a bit more complicated. We solve this problem in full generality and show that the quasinormal modes defined are the zeroes of the determinant spanned by the values at the boundary of a maximal set of linearly independent solutions to the field equations.

Having solved the problem of defining the holographic Green functions we concentrate on finding the lowest quasinormal modes and in particular the ones representing the hydrodynamic behavior of the system. Hydrodynamic modes can be understood as massless modes in the sense that $\lim_{k\rightarrow 0}\omega(k)=0$. Such modes arise in the presence of a conserved charge. In this case a local charge distribution can not simply dissipate away but has to spread slowly over the medium  according to a diffusion process. Other situations in which hydrodynamic modes appear are at a second order phase transition, characterized by the appearance of a new massless mode and spontaneous breaking of a global symmetry where a massless Goldstone boson appears. A discussion of hydrodynamics in systems with spontaneous breaking of global symmetries can be found in \cite{ChaikinLubensky} and in the relativistic context in \cite{Son:2000ht}.

We will consider the abelian gauge model of \cite{Hartnoll:2008vx} without backreaction, i.e. assuming that the metric is a simple 
AdS black hole with flat horizon topology. In \cite{Hartnoll:2008vx} it has been established that this model undergoes a second order
phase transition towards forming a condensate of the charged scalar field thus spontaneously breaking the $U(1)$ gauge
symmetry.  The conductivity in the broken phase has a delta function peak at zero frequency and a gap typical of superconductors.
We add here that although this model is referred to as holographic superconductor it is more proper to speak of a holographic
charged superfluid as in \cite{Herzog:2008he} since the $U(1)$ symmetry in the boundary field theory is global and there is no clear holographic description of how to add gauge fields. 

An important particularity of the model is the missing backreaction. Since the metric fluctuations are set to zero this means that
effectively there is no energy momentum tensor in the field theory dual. In particular the generators of translations and rotations
are missing in the operator algebra. This does of course not mean that the model does not have these symmetries, they
are however not realized as inner automorphisms of the operator algebra (they are still outer automorphisms). Besides 
the space time symmetries being realized as outer automorphisms there is a direct consequence of this in what concerns
the hydrodynamics of the model: all hydrodynamic modes related to them are missing. There is no shear mode for the
momentum diffusion and no sound mode for the energy transport. The hydrodynamic modes we find in the model are therefore
only due to the presence of the $U(1)$ symmetry and its spontaneous breakdown. 

Hydrodynamics can be understood as an effective field theory defined by the continuity equations of the conserved currents and
so-called constitutive relations which encode the dissipative behavior of the system. The constitutive relations tell us how fast a current is built up due to gradients in the charge density or due to external fields. The constitutive relations depend on transport coefficients such as viscosity or conductivity. Transport coefficients can be divided into absorptive and reactive ones depending on
whether they are odd or even under time reversal  \cite{ChaikinLubensky}. A typical example for an absorptive transport coefficient
is the diffusion constant, an example for a reactive one is the speed of sound. Reactive transport coefficients such as the speed
of sound (the static susceptibility is another example) can often be computed from purely thermodynamic considerations. This
has been done for the speed of second sound in this model in \cite{Herzog:2008he} and for a variant of fourth sound in
\cite{Yarom:2009uq}\footnote{Note that in \cite{Herzog:2008he} the mode has been called second sound whereas in \cite{Yarom:2009uq} it was argued that
it should be rather called fourth sound. There it has been argued that it is the fourth sound that survives the
probe limit. In any case working directly in the probe limit we only have one sound mode. For convenience we will
refer to it as second sound. Disentangling first, second or fourth sound would need to take into account the
backreaction which is beyond the scope of this paper.}. As we have argued before, the hydrodynamic modes of
energy and momentum transport, shear and sound modes are missing. In the broken phase one expects however
the appearance of a hydrodynamic mode with approximately linear dispersion relation for small momenta which represents
the second sound present in superfluids. Indeed such a mode is bound to appear for each spontaneously broken continuous symmetry
\cite{ChaikinLubensky}. Our aim is to find the second sound mode directly as a pole in the holographic Green functions in
the broken phase and to read off the speed of sound from its dispersion relation. We have found this mode numerically and
our results for the speed of second sound agree (with numerical uncertainties) with the results in \cite{Herzog:2008he}.

Below the critical temperature  one expects actually  only a part of the medium to be in the superfluid
state whereas another part stays in the normal fluid phase. The fluid is a two component fluid and naively one
might expect that this is reflected in the pole structure of the Green functions as the presence of the diffusive pole 
of the normal fluid component. As we will see, the hydrodynamic character of this diffusive pole is lost however
below the critical temperature. We find a pole with purely imaginary frequencies obeying a dispersion relation  
roughly of the form $\omega = -i\gamma - iD k^2$. The gap $\gamma$ goes to zero at the $T_c$ such that at
the critical temperature this mode goes over into the usual diffusive mode of the normal fluid.

Our goal in the following is to establish the presence of the second sound pole in the holographic superconductor model \cite{Hartnoll:2008vx}.
In section two we introduce the model and describe its properties. We compute the condensate as a function of temperature.
This is basically a review of the results of \cite{Hartnoll:2008vx} except the fact that we choose to work in the grand canonical ensemble where we hold fixed
the value of the chemical potential instead of the value of the charge density.

In section three we compute the quasinormal modes of the complex scalar field in the unbroken phase.
As expected we find that at the critical
temperature a quasinormal frequency crosses over into the upper half of the complex frequency plane. Since a pole
in the upper half plane is interpreted as an instability this is an indication that the scalar field condenses. The quasinormal
modes of gauge fields in the four dimensional AdS black hole have been studied before \cite{Miranda:2008vb}. In particular the longitudinal 
gauge field channel shows a diffusion pole with diffusion constant $D=3/(4\pi T)$. The holographic Green functions for gauge
fields are often calculated in a formalism that employs gauge invariant variables, i.e. the electric field strengths. For reasons
explained  in section four we prefer however to work directly with the gauge fields. The longitudinal components obey two coupled
differential equations subject to a constraint and we show in full generality how to compute the holographic Green function in
such a situation. The quasinormal mode condition boils down to setting a determinant of field values at the boundary of AdS to
zero. We find that this determinant is proportional to the electric field strength exemplifying that the poles of holographic
Green functions are gauge invariant as expected on general grounds.

Section four is the core of our paper. Here we study the lowest modes of the quasinormal mode spectrum in the superfluid phase.
We find that the longitudinal gauge fields at non zero momentum couple to real and imaginary part of the scalar field fluctuations, building up a system of four coupled differential equations subject to one constraint. We solve this system numerically and
compute the quasinormal modes from our determinant condition. We find hydrodynamic modes with approximately linear
real part of the dispersion relation. We compute the speed of second sound from it and find our results to be in good numerical agreement
with what was found in \cite{Herzog:2008he} from thermodynamic considerations. The second sound pole has however also an imaginary part and
we can fit the dispersion relation (for small momenta) to $\omega = \pm v_s k - i \Gamma_s k^2$ which allows us to read off
the attenuation constant $\Gamma_s$ of second sound. We also find a purely imaginary mode with a dispersion relation of the
form $\omega = -i \gamma - i D k^2$  with $D$ the diffusion constant. It is a sort of gapped diffusion mode. The gap $\gamma$
goes to zero for $T\rightarrow T_c$. A simple two fluid model suggests that there is still a normal fluid component and in it charges should diffuse in the usual way right below $T_c$. Therefore we expect a diffusive pole to show up in the two-point function. The diffusive behavior
is modified however at long wavelength by the presence of the gap $\gamma$. This mode is therefore not really a hydrodynamic
mode. 

We close this work with section five where we summarize and discuss our results.

%%%%%%%%%%%%%%%%%%%%%%%%%%%%%%%%%%%%%%
%%%%%%%%%%%% THE MODEL %%%%%%%%%%%%
%%%%%%%%%%%%%%%%%%%%%%%%%%%%%%%%%%%%%%
\section{\label{sec:background}The Model}
As in \cite{Hartnoll:2008vx} we consider a four dimensional planar AdS black hole with line element
\begin{equation}
\label{eq:metric}
ds^2 = - f(r) dt^2 + \frac{dr^2}{f(r)} + \frac{r^2}{L^2} (dx^2+dy^2)\,.
\end{equation}
the blackening factor is $f(r) = \frac{r^2}{L^2}-\frac{M}{r}$.  This metric has  a horizon at $r_H = M^{1/3}L^{2/3}$, the 
Hawking temperature is $T= \frac{3}{4\pi} \frac{r_H}{L^2}$. In the following we will rescale coordinates according to
\begin{equation}
\left( \begin{array}{c}
r\\
t\\
x\\
y
\end{array}
\right)
\rightarrow 
\left( \begin{array}{c}
r_H \rho\\
L^2/r_H\, \bar t\\
L^2/r_H\, \bar x\\
L^2/r_H\, \bar y
\end{array}
\right)
\end{equation}
In the new dimensionless coordinates the metric takes the form \eqref{eq:metric} with $M=1$ times the overall AdS scale $L^2$. 

We take an abelian gauge model with a massive charged scalar field
\begin{equation}
\label{eq:abelianHiggs}
\cL = -\frac 1 4 F_{\mu\nu} F^{\mu\nu}  - m^2 \Psi \bar\Psi - (\partial_\mu \Psi - i A_\mu \Psi)(\partial^\mu \bar \Psi + i A^\mu \bar\Psi)
\end{equation}
and a tachyonic mass $m^2=-2/L^2$ above the Breitenlohner-Friedmann bound.
As in \cite{Hartnoll:2008vx} we ignore the backreaction of these fields onto the metric. We seek solutions for which the time component
of the gauge field vanishes at the horizon and takes a non-zero value  $\mu$ on the boundary. This value can be
interpreted as the chemical potential. The boundary condition on the horizon is usually justified by demanding that
the gauge field has finite norm there. Here this can be seen as follows: as in \cite{Herzog:2008he} we can chose a 
gauge that removes the phase of the scalar field $\Psi$ from the equations of motion. In this gauge the scalar field
current becomes $J_\mu = \psi^2 A_\mu$. Therefore the value of $A_\mu$ is directly related to a physical quantity,
the current, and it is a well defined physical condition to demand the current to have finite norm at the horizon.
This is achieved by taking the scalar field to be regular and the gauge field to vanish at the horizon.

 In addition to the gauge field the scalar field might be non trivial as well. In fact
for high chemical potential (low temperature) the scalar field needs to be switched on in order to have a stable
solution. In our study of the quasinormal modes in the next section we will indeed see that a quasinormal mode
crosses into the upper half plane at the critical temperature. We denote the temporal component of the gauge field
in the dimensionless coordinates by $\Phi$. The field equations for the background fields are

\begin{eqnarray}\label{eq:background}
\Psi'' +(\frac{f'}{f} + \frac 2 \rho) \Psi' + \frac{\Phi^2}{f^2} \Psi + \frac{2}{L^2 f} \Psi &=& 0 \,,\\
\Phi'' +  \frac 2 \rho \Phi' - \frac{2 \Psi^2}{f} \Phi &=&0\,.
\end{eqnarray}
The equations can be solved numerically by integration from the horizon out to the boundary. As we just argued
for the current to have finite norm at the horizon we have to chose
$\Phi(1)=0$ and demand the scalar field to be regular at the horizon. These conditions leave two integration
constants undetermined. The behavior of the fields at the conformal boundary is
{
\begin{eqnarray}\label{eq:bgboundary1}
\Phi &=& \bar\mu - \frac{\bar n}{\rho} + O(\frac{1}{\rho^2}) \,,\\\label{eq:bgboundary2}
\Psi &=& \frac{\psi_1}{\rho} + \frac{\psi_2}{\rho^2}  + O(\frac{1}{\rho^2}) \,.
\end{eqnarray}}
The value of the mass of the scalar field chosen allows to define two different theories due to
the fact that both terms in the expansion above are normalizable in AdS. 
The canonical choice of what one considers to be the normalizable mode gives a theory in which $\psi_1$ is 
interpreted as a coupling and $\psi_2$ as expectation value of an operator
 with mass dimension two. On the other hand one might consider $\psi_2$ as the coupling and $\psi_1$ as
the expectation value of an operator of dimension one.

All our numerical calculations are done using the dimensionless coordinates. In order to 
relate the asymptotic values \eqref{eq:bgboundary1}, \eqref{eq:bgboundary2} to the physical quantities we note that
\begin{eqnarray}
\label{eq:physicalquantities}
\bar \mu &=& \frac{3 L}{4 \pi T} \mu\,,\\
\bar n &=& \frac{9 L}{16 \pi^2 T^2} n\,,\\
\psi_1 &=& \frac{3}{4 \pi T L^2} \langle O_1 \rangle\,,\\
\psi_2 &=& \frac{9}{16 \pi^2 T^2 L^4} \langle O_2 \rangle\,,
\end{eqnarray} 
where $\mu$ is the chemical potential, $n$ the charge density and $\langle O_i\rangle$ are the vacuum expectation
values of the operators sourced by the scalar field. From now on we will set $L=1$ and work in the grand canonical
ensemble by fixing $\mu =1$. Different values for $\bar\mu$ can now be interpreted as varying the temperature $T$.
For high temperatures the scalar field is trivial and the gauge field equation is solved by $\Phi = \bar \mu - \frac{\bar \mu}{\rho}$. Spontaneous symmetry breaking means that an operator has a non trivial expectation value
even when no source for the operator is switched on. We therefore look for nontrivial solutions of the scalar field
with either $\psi_1=0$ or $\psi_2=0$.
Numerically we find that a non-trivial scalar field is switched on at $\bar \mu = 1.1204$ corresponding to a critical temperature $T_c = 0.213 \mu$ for the operator $O_1$ and at $\bar \mu = 4.0637$ corresponding to a critical temperature of $T_c = 0.0587 \mu$ for the operator $O_2$. 
\begin{figure}[!htbp]
\centering
\begin{tabular}{cc}
\includegraphics[width=7cm]{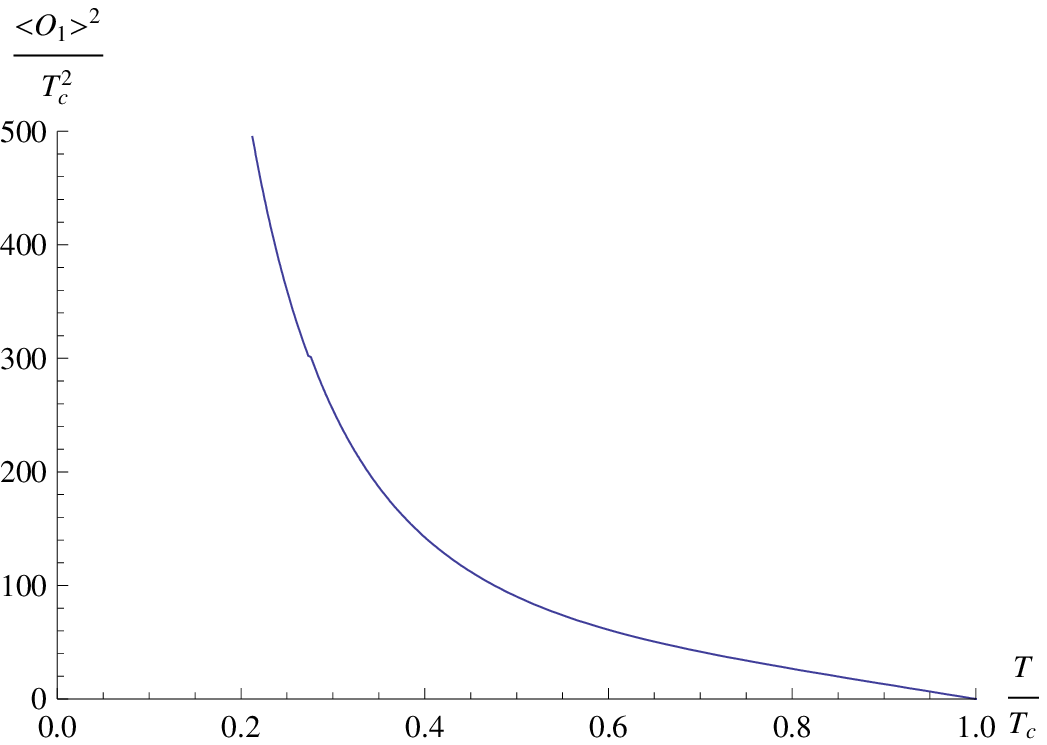} & \includegraphics[width=7cm]{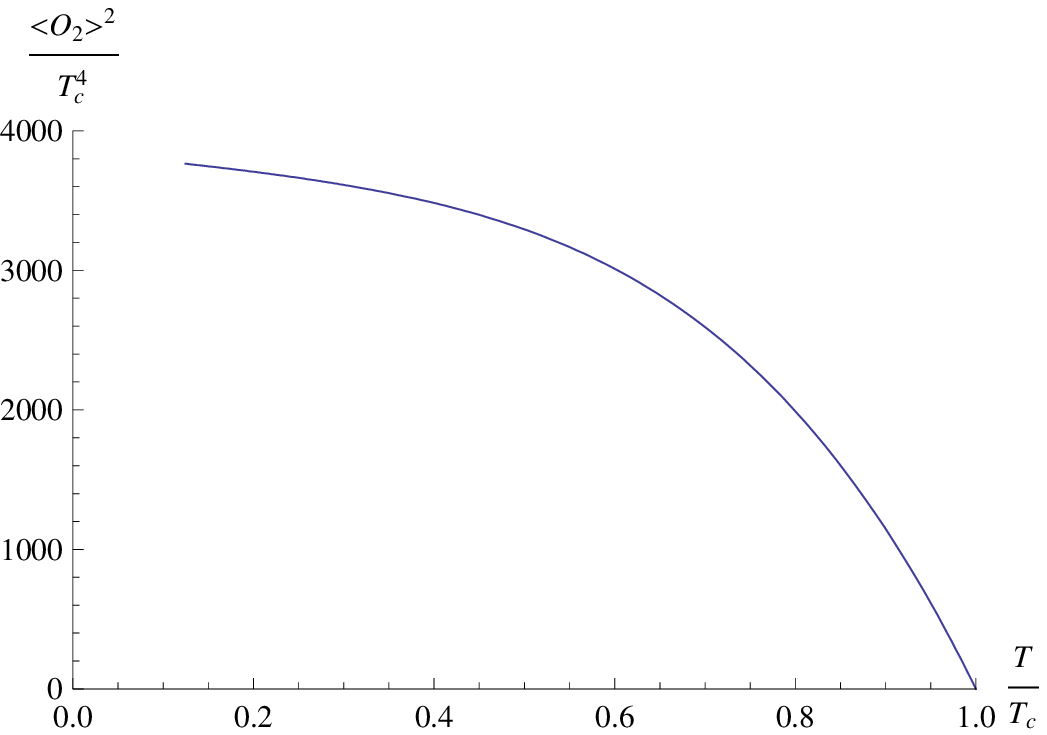} 
\end{tabular}
\caption{\label{fig:condensates} The condensates as function of the temperature in the two possible theories.}
\end{figure}
We chose to plot the squares of the condensates as a function of reduced temperature. It makes the linear
behavior for temperatures just below the critical one manifest. 
\begin{equation}
\langle O_i \rangle^2 \propto \left(1-\frac{T}{T_c} \right)\,.
\end{equation}

Since we want to compute the holographic two point functions we will have to expand the action to second order
in field fluctuations around the background. We divide the fields into background plus fluctuations in
the following way
\begin{eqnarray}\label{eq:flucs}
\Psi &=& \psi(\rho) + \sigma(\rho,t,x) + i \eta(\rho,t,x)\,,\\
A_\mu &=& \cA_m(\rho) +  a_\mu(\rho,t,x)\,.
\end{eqnarray}
The gauge transformations act only on the fluctuations 
\begin{eqnarray}\label{eq:gaugetrafos}
\delta a_\mu &=&\partial_\mu \lambda\,,\\
\delta \sigma &=& -\lambda \eta\,,\\
\delta \eta &=& \lambda \sigma + \lambda \psi\,.
\end{eqnarray}
The action expanded out to second order is $S= S^{(0)} + S^{(1)} + S^{(2)}$
\begin{eqnarray}\label{eq:s0}
S^{(0)} &=& \int\, \sqrt{-g} \left(
-\frac{1}{4} \cF_{\mu\nu} \cF^{\mu\nu}  + \frac{2}{L^2} \psi^2 - (\partial_\mu\psi) (\partial^\mu\psi) - \cA_\mu \cA^\mu \psi^2 \right) \,,\nonumber\\
S^{(1)} &=&  \int\, \sqrt{-g} \left( -\frac 1 2 \cF_{\mu\nu} f^{\mu\nu} + \frac 4 L^2 \psi \sigma - 2 \partial_\mu\psi
\partial^\mu \sigma - 2 \partial_\psi \cA^\mu \eta - 2 \cA^2 \psi \sigma + 2 \cA_\mu\partial^\mu\eta - 2 \cA_\mu a^\mu \psi^2 \right)\,,\nonumber\\
S^{(2)} &=&  \int\, \sqrt{-g} \left( -\frac 1 4 f_{\mu\nu} f^{\mu\nu} - (\partial \sigma)^2 - (\partial \eta)^2 - \cA^2 \sigma^2 - \cA^2\eta^2 + \frac 2 L^2 \sigma^2 +\frac 2 L^2 \eta^2 - 2 \partial_\mu\psi a^\mu \eta \right. \nonumber\\
& & \left. - 2 \cA^\mu a_\mu \psi \sigma -
2 \partial_\mu\sigma \cA^\mu \eta + 2 \cA_\mu \sigma\partial\mu\eta + 2 \partial_\mu \eta a^\mu \psi - a^2\psi^2
\vphantom{\frac{1}{1}}\right)\,.
\end{eqnarray}
Up to the equations of motion  these can be written as boundary terms
\begin{eqnarray}\label{eq:boundaryaction}
S^{(1)}_B &=& \int_B \sqrt{-g} \left( g^{\rho\rho} \cF_{\rho\mu} a^\mu - 2 g^{\rho\rho} \partial_\rho \psi \sigma + 2 A^\rho \psi \eta \right)\,,\nonumber\\
S^{(2)}_B &=& -\int_B \sqrt{-g} g^{\rho\rho}\left( \frac 1 2 g^{\nu\lambda}f_{\rho\nu}a_\lambda  + \eta \partial_\rho \eta + \sigma \partial_\rho \sigma + a_\rho \psi \eta \right)\,.
\end{eqnarray}
Note that $S^{(1)}$ is not trivial since according to the holographic dictionary it has to encode the non-vanishing
expectation values of the field theory operators.

%%%%%%%%%%%%%%%%%%%%%%%%%%%%%%%%%%%%%%
%%%%%%%%%%%% UNBROKEN PHASE  %%%%%%%%%%%%
%%%%%%%%%%%%%%%%%%%%%%%%%%%%%%%%%%%%%%
\section{\label{sec:unbroken}Quasinormal Frequencies in the Unbroken Phase}

We assume now that the scalar field $\Psi$ has vanishing background value and take the field to depend
on $t,x,\rho$. Frequency $\omega$ and momentum $k$ in the dimensionless coordinates are related to the physical ones
$\omega_{ph}$, $k_{ph}$ as
\begin{equation}\label{eq:freqmom}
\omega = \frac{ 3 \omega_{ph}}{4 \pi T} ~~~,~~~ k = \frac{ 3  k_{ph}}{4 \pi T}\,.
\end{equation}
The equations of motion in the unbroken phase are
\begin{eqnarray}\label{eq:eomsnormal}
0 &=&\Psi'' +(\frac{f'}{f}+\frac{2}{\rho} )\Psi' + (\frac{(\Phi+\omega)^2}{f^2} +\frac{2}{f} - \frac{k^2}{f \rho^2})\Psi\,,\nonumber\\
0 &=&  a_t'' +\frac{2}{\rho} a_t' - \frac{k^2}{\rho^2}a_t - \frac{\omega k}{f \rho^2} a_x\,,\nonumber\\
0 &=&  a_x'' + \frac{f'}{f} a_x' +\frac{\omega^2}{f^2} a_x + \frac{\omega k}{f \rho^2} a_t \,,\nonumber\\
0 &=& \frac{\omega}{f} a_t' + \frac{k}{\rho^2} a_x'\,.
\end{eqnarray}
The equation of motion for the complex conjugate scalar $\bar\Psi$ can be obtained by changing the sign of the
gauge field background $\Phi$ in \eqref{eq:eomsnormal}.

\subsection{Green Functions}\label{ssec:Greenfs}
In order to calculate the quasinormal frequencies we impose ingoing boundary conditions at the horizon. Since the coefficients
of the differential equations \eqref{eq:eomsnormal} are known analytically and are such that they are of Fuchsian type we can
use the Frobenius method to approximate the solutions at the horizon and at the boundary by series expansions. As is well-known
the holographic Green functions are proportional to the ratio of the connection coefficients. More precisely we demand
\begin{equation}
\Psi_H = (\rho-1)^{-i\omega/3} ( 1+ O(\rho-1) )\,,
\end{equation}
on the horizon and write the local solution at the AdS boundary as
\begin{equation}\label{eq:bdryexpansion}
\Psi_B = \frac{A}{\rho} + \frac{B}{\rho^2} + \O\left(\frac{1}{\rho^3}\right)\,.
\end{equation}
In the theory with the dimension two operator we take $A$ as  the coefficient of the non-normalizable mode and $B$ as the coefficient of the normalizable mode.
Writing the local solution on the horizon as a linear combination of normalizable and non-normalizable
modes on the boundary fixes the connection coefficient $A$ and $B$. We have written the boundary action
as a functional of real and imaginary part of the scalar field. 
We will rewrite the boundary action now in terms of the complex scalar
$\Psi$ and its conjugate. In addition we introduce a local boundary counterterm to regularize
the action.
\begin{equation}
S^{B}_\Psi =\int \left[ -\frac 1 2 f \rho^2 (\bar\Psi \Psi'+ \Psi \bar\Psi') -\rho^3 \bar\Psi \Psi\right]_{\rho=\Lambda}\,. 
\end{equation}
This allows to compute the Green functions $G_{\bar O O}(q)=\langle \bar O(-q)  O(q)\rangle$ and 
$G_{O\bar O }(q)=\langle O(-q) \bar O(q) \rangle$ fulfilling $G_{O\bar O}(-q)= G_{\bar O O}(q)$,  with $q$ being the four momentum $(\omega,k)$.
We write $\Psi(q,\rho) = \Psi_0(q) f_q(\rho)$, where we interpret $\Psi_0(q)$ as the source that
inserts the operator $O(q)$ in the dual field theory. We introduce a cutoff and normalize the profile
function $f_q$ to $1/\Lambda$ at the cutoff. In terms of an arbitrary solution the normalized one is  $f_k(\rho) = \Psi_k(\rho)/(\Lambda\Psi(\Lambda))$ where $\Psi_k(\rho)$ has the boundary expansion (\ref{eq:bdryexpansion}). 
The boundary action is now
\begin{eqnarray}
S^B &=& -\frac 1 2 \int \left[\Psi_0(-q)(\rho^2 f f_{-q} \bar f'_q + \rho^3 f_{q} \bar f_q) \bar \Psi_0(q) + \bar\Psi_0(-q)(\rho^2 f \bar f_{-q} f'_q + \rho^3 \bar f_{-q} f_q)  \Psi_0(q)\right]_{\rho=\Lambda }\nonumber\\ 
&=&\int \Psi_0(-q)    \cF_{\Psi\bar\Psi}(\Lambda) \bar \Psi_0(q) +\bar \Psi_0(-q)\cF_{\bar\Psi\Psi}(\Lambda)\Psi_0(q)\,.
\end{eqnarray}
According to the holographic dictionary the renormalized retarded Green functions are given by
the limit $\lim_{\Lambda\rightarrow\infty} -2 \cF(\Lambda)$:
\begin{equation}\label{eq:unbrokenGFsO2}
G_{\bar O_2 O_2} =   \frac{B}{A} ~~~,~~~ G_{O_2 \bar O_2} =   \frac{\bar B}{\bar A}\,.
\end{equation}
We denote the connection coefficients for the complex conjugate scalar as $\bar A$, $\bar B$\footnote{Note that
the infalling boundary condition for the conjugate scalar is  $\bar \Psi \sim (\rho-1)^{- i \omega/3}$.}. According to \eqref{eq:eomsnormal}
they can be obtained by switching the sign of the chemical potential in the expressions for $A,B$.

The theory with the operator of dimension one can be obtained through a Legendre transform. We note that 
the expectation value of the dimension two operator is $\langle O_2 \rangle = -\left. \rho^2 (\rho\Psi(\rho))'\right|_{\rho=\Lambda}$,
whereas the source is given by $ \Lambda \Psi(\Lambda)$.
We therefore add the following terms to the boundary action
\begin{equation}
S^B  \rightarrow S^B + \int d^4k \left.\left[ \rho^3 (\bar\Psi(\rho\Psi(\rho))'  + 
\Psi(\rho\bar\Psi(\rho))' \right]\right|_{\rho=\Lambda}\,.
\end{equation} 
Now we can evaluate the Green functions as before, the only difference being the normalization of the profile
function $f_k(\rho) = \Psi_k(\rho)/[ \Lambda \Psi'(\Lambda) + \Psi(\Lambda) ]$. 
This normalization takes care that the term of order $1/\rho^2$ in the boundary expansion couples
with unit strength to the source $\Psi_0(q)$.
We find finally for the Green functions of the Legendre transformed theory
\begin{equation}\label{eq:unbrokenGFsO1}
G_{O_1 \bar O_1} =  \frac{A}{B} ~~~,~~~ G_{\bar O_1 O_1} =   \frac{\bar A}{\bar B}\,,
\end{equation}
as expected. The quasinormal modes in the scalar sector are given by the zeroes of the connection coefficients
$A$ and $\bar A$ in the theory with operator of dimension two and by the zeroes of $B$ and $\bar B$ in the 
theory with the dimension one operator. In terms of the unnormalized solutions to the field equations we can write the Green 
functions as
\begin{eqnarray}
G_{\bar O_2 O_2} &=& - \lim_{\Lambda\rightarrow \infty} \left(\Lambda^2\frac{\Psi'_q(\Lambda)}{\Psi_q(\Lambda)} +\Lambda\right)
\,,\\
G_{O_1\bar O_1} &=&  \lim_{\Lambda\rightarrow \infty}  \frac{\Psi_q(\Lambda)}{\Lambda(\Lambda\Psi'_q(\Lambda)+\Psi_q(\Lambda))} \,.
\end{eqnarray}

\subsection{Quasinormal Modes from Determinants}\label{ssec:detmethod}
Before presenting the results for the quasinormal modes of the scalar field we would like to outline a method of how to
calculate the holographic Green functions for the gauge fields without using gauge invariant variables such as the electric
field strength $E= -i( k a_t + \omega a_x)$.

The complicated structure of the gauge symmetry in the broken phase makes it rather difficult to
express the boundary action in terms of gauge invariant field combinations. As a warm up for the
problem of how to calculate the holographic Green functions in this situation we will consider how
we can calculate them in the unbroken phase directly in terms of the gauge fields.
We necessarily have to solve a system of coupled
differential equations whose solutions are restricted by a constraint.

The correct boundary conditions for the gauge fields on the horizon are
\begin{eqnarray}\label{eq:bcinfalling}
a_t &\propto& (\rho-1)^{1-i  \omega/3} ( a_t^0 + \dots) \,,\\
a_x &\propto& (\rho-1)^{-i  \omega/3} ( a_x^0 + \dots) \,.
\end{eqnarray}
The two coefficients $a^0_x$ and $a^0_t$  are not independent but related by the constraint.
At this point we have fixed the incoming wave boundary conditions and there seems to be now a unique solution
to the field equations. We would expect however two linearly independent solutions with infalling boundary conditions on the
horizon. The constraint reduces this to only one solution.
The problem is now that in order to compute the Green function for the charge density and longitudinal current
component separately we need solutions that asymptote to $(a_t,a_x) = (1,0)$ and $(a_t, a_x)  =(0,1)$ respectively.
This is of course not possible with only one available solution at the horizon.
Because of the gauge symmetry  the gauge field system (\ref{eq:eomsnormal}) allows for an algebraic solution
\begin{eqnarray}\label{eq:zeromode}
a_t &=& - \omega \lambda \,,\\
a_t &=& k \lambda  \,,
\end{eqnarray}
with $\lambda'=0$, i.e. $\lambda$ being independent of $\rho$. This is of course nothing but a gauge transformation of
the trivial solution. Remember that
even after fixing the radial gauge $a_\rho=0$, gauge transformations with gauge parameters independent of $\rho$ are still possible.
These gauge transformations appear as algebraic solutions to the field equations. We also stress that the infalling boundary
conditions really have to be imposed only on physical fields, i.e. the electric field strength. Having therefore an arbitrary
non trivial gauge field solution corresponding to an electric field with infalling boundary conditions we can add to it
the gauge mode \eqref{eq:zeromode}.

We can use this to construct a basis of solutions that allows the calculation of the holographic Green functions.
Let us now assume that there is a solution that takes the values $(a_t,a_x) = (1,0)$ at the boundary. We will call this
solution from now on $\alpha^t_i$ for $i\in{t,x}$. Analogously we define the solution $\alpha^x_i$. According to the holographic
dictionary the solution $\alpha^t_i$ couples to the boundary value $\lim_{\rho\rightarrow \infty} a_t(q,\rho)= \cA_t(q)$, i.e. the source of the field
theory operator $J_t$ (the time component of the conserved current $J_\mu$). In parallel $\alpha_i^x$ couples to the boundary
value $\cA_x(q)$. A generic solution of the gauge field equations can now be written in terms of
the boundary fields as
\begin{equation}
a_i(q,\rho) = \cA_x(q) \alpha^x_i(\rho) + \cA_t(q) \alpha^t_i(\rho)\,.
\end{equation}
Using this expansion the boundary action can be written as
\begin{equation}
S_B =  \frac 1 2 \int_B \cA_i(-q)\left[   \left(\rho^2 \alpha^i_t(-q,\rho) \frac{d}{d\rho} (\alpha^j_t(q,\rho)) -
f(\rho)   \alpha^i_x(-q,\rho) \frac{d}{d\rho} (\alpha^j_x(q,\rho)) \right)  \right]_{\rho=\Lambda} \cA_j(q)\,,
\end{equation}
where again we have introduced a cutoff at $\rho = \Lambda$. From this it follows that the holographic Green functions are given by
\begin{eqnarray}
2\cF^{ij}(\rho) &=&   \rho^2 \alpha^i_t(-k,\rho) \frac{d}{d\rho} (\alpha^j_t(k,\rho)) -
f(\rho)   \alpha^i_x(-k,\rho) \frac{d}{d\rho} (\alpha^j_x(k,\rho)) \,,
\end{eqnarray}
in the limit
\begin{equation}\label{eq:Greenfns}
G^{ij} = \lim_{\Lambda\rightarrow \infty}-2 \cF^{ij}(\Lambda)\,.
\end{equation}
Notice also that $\frac{d}{d\rho} [\cF^{ij}(\rho)-\cF^{*ji}(\rho)]=0 $ by the field equations.

Although there are no terms of the form $a_t' a_x$ in the boundary action this formalism gives automatically expressions for the mixed Green functions $G_{tx}$ and $G_{xt}$!
But we still have to construct the solutions $\alpha^i_j(\rho)$. This can be done in the following way: suppose we have
an arbitrary solution $(a_t(\rho), a_x(\rho))$ obeying the infalling boundary conditions \eqref{eq:bcinfalling}.
We can add to this now an appropriate
gauge mode, such that at the cutoff the solution takes the form $(1,0)$ or $(0,1)$ in terms of $a_t(\Lambda), a_x(\Lambda)$.
This is easily achieved by solving
\begin{equation} \label{eq:linearsystem}
\left( \begin{array}{cc}
c_t^t & c_x^t\\
c_t^x & c_x^x
\end{array}
\right)
\left(
\begin{array}{cc}
a_t(\Lambda) & a_z(\Lambda)\\
-\omega \lambda & k \lambda
\end{array}
\right)
=
\left( \begin{array}{cc}
1 & 0\\
0 & 1
\end{array}
\right)\, .
\end{equation}
The linear combinations formed with the coefficients $c^i_j$ give now new solutions $(\alpha^t_t(\rho), \alpha^t_x(\rho) )$ and  $(\alpha^x_t(\rho), \alpha^x_x(\rho) ) )$ obeying the correct boundary conditions on the AdS boundary.
Using the general expression for the Green function \eqref{eq:Greenfns} we get explicitly in terms of solutions obeying the
infalling boundary conditions
\begin{eqnarray}
G_{tt} &=& \lim_{\Lambda \rightarrow \infty} \Lambda^2 \frac{ k a'_t(\Lambda)}{ k a_t(\Lambda) + \omega a_x(\Lambda)}\,,\\
G_{tx} &=& \lim_{\Lambda \rightarrow \infty} \Lambda^2 \frac{ \omega a'_t(\Lambda)}{ k a_t(\Lambda) + \omega a_x(\Lambda)}\,,\\
G_{xt} &=& -\lim_{\Lambda \rightarrow \infty} \Lambda^2 \frac{ k a'_x(\Lambda)}{ k a_t(\Lambda) + \omega a_x(\Lambda)}\,,\\
G_{xx} &=& -\lim_{\Lambda \rightarrow \infty} \Lambda^2 \frac{ \omega a'_x(\Lambda)}{ k a_t(\Lambda) + \omega a_x(\Lambda)}\,.
\end{eqnarray}
Note that the denominator for all is given by $ k a_t(\Lambda) + \omega a_x(\Lambda)$ which
is up to an irrelevant constant nothing but the gauge invariant electric field $E_x$.
Therefore we see immediately that the poles of these Green function
are gauge invariant and coincide of course with the poles of the Green function in the gauge invariant formalism where
$
G\propto \frac{E_x'}{E_x} 
$.
Indeed using the constraint on the boundary we find the well known expressions \cite{Kovtun:2005ev}
\begin{eqnarray}
G_{tt} &=& \frac{k^2}{k^2-\omega^2} \lim_{\Lambda \rightarrow \infty} \Lambda^2  \frac{E_x'}{E_x}  ~~~,~~~
 G_{tx} = \frac{k \omega}{k^2-\omega^2} \lim_{\Lambda \rightarrow \infty} \Lambda^2  \frac{E_x'}{E_x}\,,\nonumber\\
G_{xx} &=& \frac{\omega^2}{k^2-\omega^2} \lim_{\Lambda \rightarrow \infty} \Lambda^2  \frac{E_x'}{E_x}\,.
\end{eqnarray}
On general grounds one expects indeed that the poles of the holographic Green functions for gauge fields are gauge independent.

If we are interested only in the location of quasinormal frequencies we do not even have to construct the holographic Green functions
explicitly. From the linear system in \eqref{eq:linearsystem} we infer that the quasinormal frequencies coincide with the zeroes
of the determinant of the field values at the boundary. Indeed vanishing determinant means that there is a nontrivial zero mode
solution to \eqref{eq:linearsystem} such that the boundary values of the fields are $(0,0)$ which in turn means that the coefficient
in the solution of the non-normalizable mode vanishes. The determinant is $\lambda( k a_t + \omega a_x)$ and again given
by the electric field strength. In fact these remarks apply to systems of coupled differential equations in AdS black hole metric in general: the quasinormal frequencies corresponding to the poles of the holographic Green functions are  the zeroes of the determinant of the field values on the boundary for a maximal set of linearly independent solutions obeying infalling boundary conditions on the horizon. The fact that the differential equations are coupled is the holographic
manifestation of mixing of operators under the RG flow. Therefore one has to specify at which scale one is defining
the operators. The scheme outlined above is dual to define the operators at the cutoff $\Lambda$.

%__________________________________________________________
\subsection{Hydrodynamic and higher QNMs}

\begin{figure}[!htbp]
\centering
\includegraphics[scale=0.65]{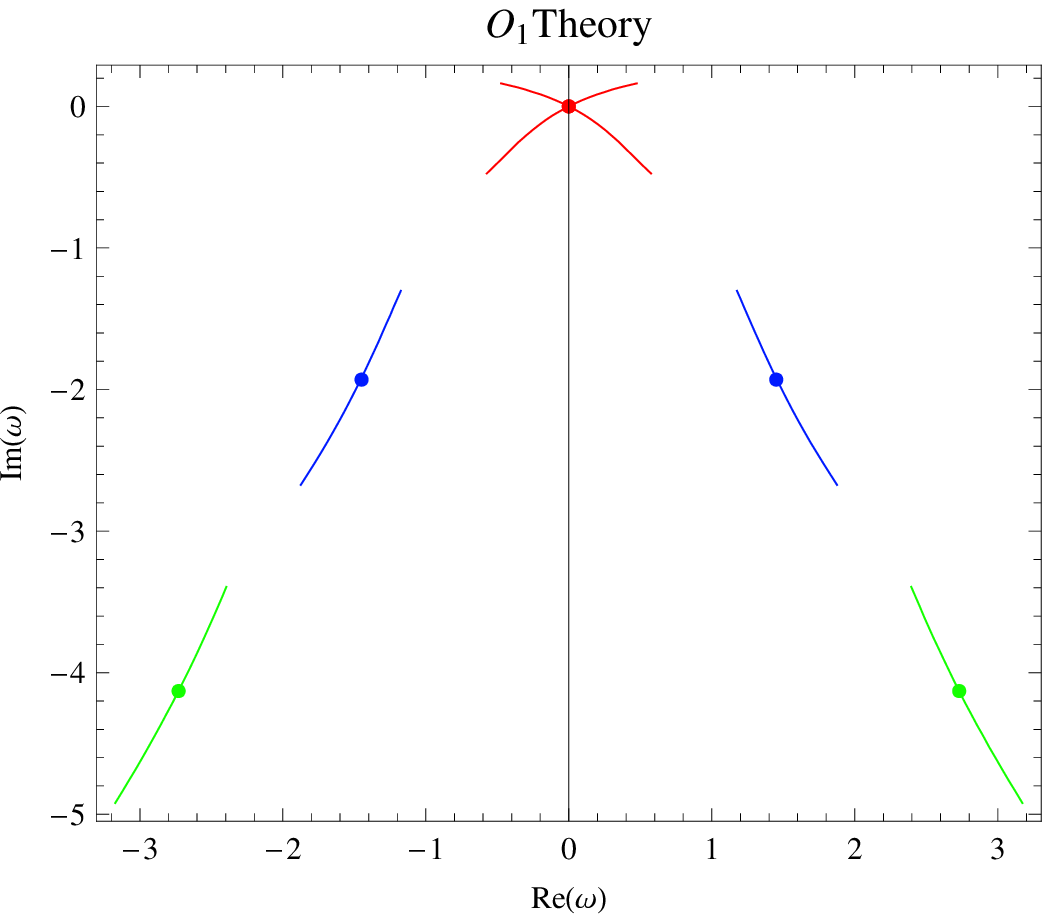} \quad \includegraphics[scale=0.65]{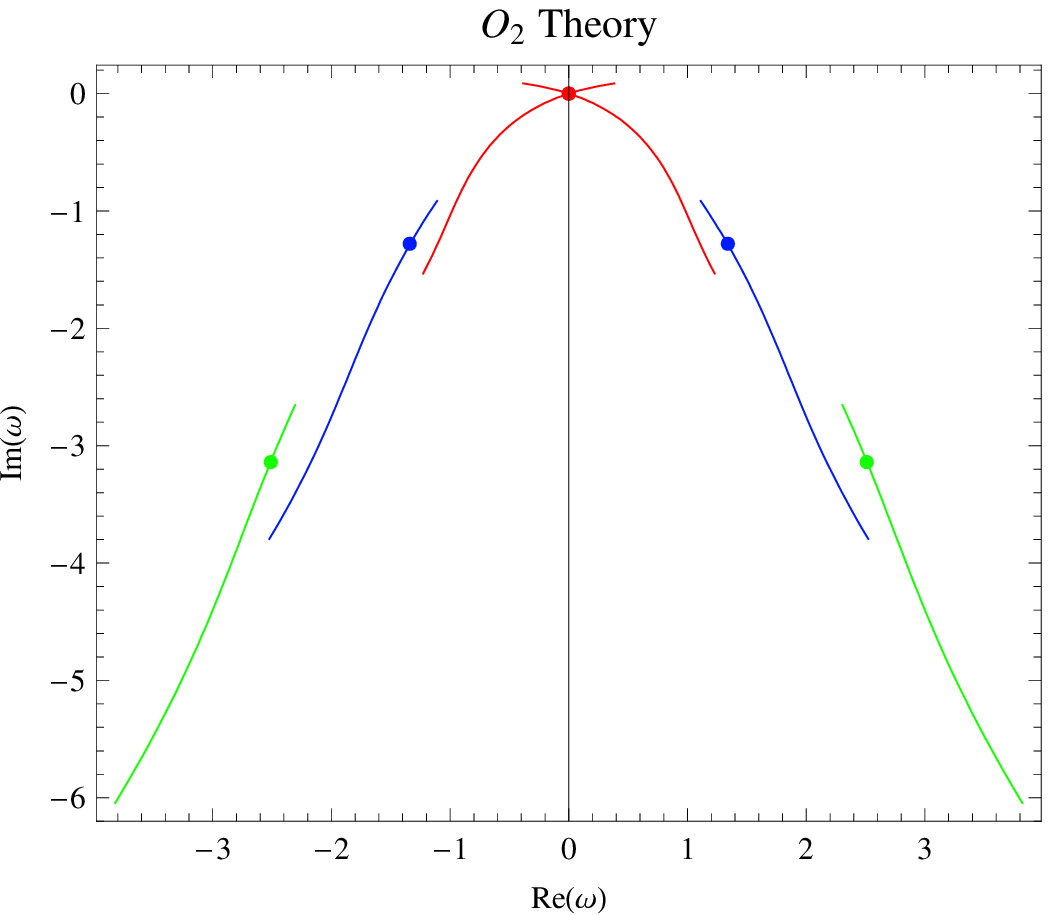}
\caption{\label{fig:scalarNP} Lowest scalar quasinormal frequencies as a function of the temperature and at momentum $k=0$, from $T/T_c=\infty$ to $T/T_c=0.81$ in the $O_2$ theory (right) and to $T/T_c=0.56$ in the $O_1$ theory (left). The dots correspond to the critical point $T/T_c=1$ where the phase transition takes place. Red, blue and green correspond to first, second and third mode respectively.}
\end{figure}

We have numerically computed the quasinormal frequencies for the fluctuations satisfying the equations of motion \eqref{eq:eomsnormal} for both the $O_2$ and the $O_1$ theories. The quasinormal modes of the scalar field correspond to zeroes of $A$ in the theory of dimension two operator and to zeroes of $B$ in the dimension one operator theory, where $A$ and $B$ are the connection coefficients of the boundary solution \eqref{eq:bdryexpansion}. Results for the lowest three poles of the scalar field at zero momentum are shown in Figure \ref{fig:scalarNP}.

The poles with positive real part correspond to the quasinormal modes of the complex scalar $\Psi$, while those with negative real part are the quasinormal modes of $\bar\Psi$, obtained by changing the overall sign of the gauge field background $\Phi$. As the temperature is decreased, the poles get closer to the real axis, until at the critical temperature $T_c$ the lowest mode crosses into the upper half of the complex frequency plane. It happens at $T_c=4.0637$ for the theory of dimension two operator and at $T_c=1.1204$ in the case of the dimension one operator theory. For $T<T_c$ the mode would become tachyonic, i. e. unstable. This instability indicates that the scalar field condenses and the system undergoes a phase transition at $T=T_c$. At the critical point, the lowest scalar quasinormal mode is a hydrodynamic mode in the sense that it is massless, lim$_{k\rightarrow0}\omega(k)=0$. This mode is identified with the Goldstone boson that appears after the spontaneous breaking of the global $U(1)$ symmetry and in the next section we will see that it evolves into the second sound mode characteristic of superfluid models. 

The quasinormal modes correspond to simple poles of the retarded Green function, so close to the $n$th pole the Green function can be approximated by $G(\omega,k,T)\sim \frac{R_n(\omega,k,T)}{\omega-\omega_n(k,T)}$. Knowing the connection coefficients we can compute the Green functions and therefore the residue for each quasinormal mode as explained in \cite{Amado:2007yr,Amado:2008ji}. For the lowest quasinormal mode at $k=0$ and at the critical temperature, the residue takes the value $R_2(T_c)=-2.545+0.825 i$ in the $O_2$ theory and $R_1(T_c)=0.686-0.348 i$ in the $O_1$ theory. In general, one expects the residues of hydrodynamic modes that correspond to conserved quantities of the system to vanish in the limit of zero momentum, since its susceptibility remains constant. Consider for instance the diffusion mode associated to conserved density. The susceptibility is defined through the two point correlation function as
\begin{equation}
 \chi=\lim_{k,\omega\rightarrow0}\langle\rho\rho\rangle=\lim_{k,\omega\rightarrow0}\frac{i\sigma k^2}{\omega+i D k^2}=\frac{\sigma}{D}\,,
\end{equation}
where $D$ is the diffusion constant and $\sigma$ is the conductivity. The residue, $i\sigma k^2$, vanishes and one recovers the well-known Einstein relation $\sigma=D\chi$. However, for hydrodynamic modes appearing at second order phase transitions the order parameter susceptibility should diverge at the critical point. This order parameter susceptibility is given in our case by the correlator of the boundary operator sourced by the scalar field. At the critical temperature it is 
\begin{equation}
 \chi_{{\bar O}_i O_i}=\lim_{k,\omega\rightarrow0}\langle{\bar O}_i O_i\rangle=\lim_{k,\omega\rightarrow0}\frac{R_i(k,T_c)}{\omega-\omega_H(k,T_c)} \rightarrow \infty
\end{equation}
since $\omega_H(0,T_c)=0$ while the residue remains finite. This result allows us to identify the lowest scalar quasinormal mode in the unbroken phase with the Goldstone boson appearing at the critical point.

In the model under consideration one can also compute the gauge field fluctuations in the normal phase. Nevertheless, as the model does not include the backreaction of the metric, the computation is not sensitive to temperature anymore. This can be seen from the equations of motion \eqref{eq:eomsnormal} of the gauge fluctuations, that do not depend on the background solutions thus are independent of the temperature. Hence we recover the results for the quasinormal modes of vector field perturbations in the AdS$_4$ black hole background computed by \cite{Miranda:2008vb}. For our purposes
the most important fact is the presence of a hydrodynamic mode corresponding to diffusion. For small momenta that
mode has dispersion relation $\omega = -i D k^2$ with $D=1$ (which is $D=3/(4\pi T)$ in physical units).
In order to study the behavior of the diffusion pole in the unbroken phase as a function of the temperature one has to consider the backreacted model described in \cite{Hartnoll:2008kx}.

%%%%%%%%%%%%%%%%%%%%%%%%%%%%%%%%%%%%%%
%%%%%%%%%%%% BROKEN PHASE %%%%%%%%%%%%
%%%%%%%%%%%%%%%%%%%%%%%%%%%%%%%%%%%%%%
\section{\label{sec:broken}Quasinormal Frequencies in the Broken Phase}

\begin{figure}[!htbp]
\psfrag{Rew}{$\text{Re}\,\omega$}
\psfrag{Imw}{$\text{Im}\,\omega$}
\psfrag{k}{$ k$}
\centering
\begin{tabular}{c}
 \includegraphics[width=0.8\textwidth]{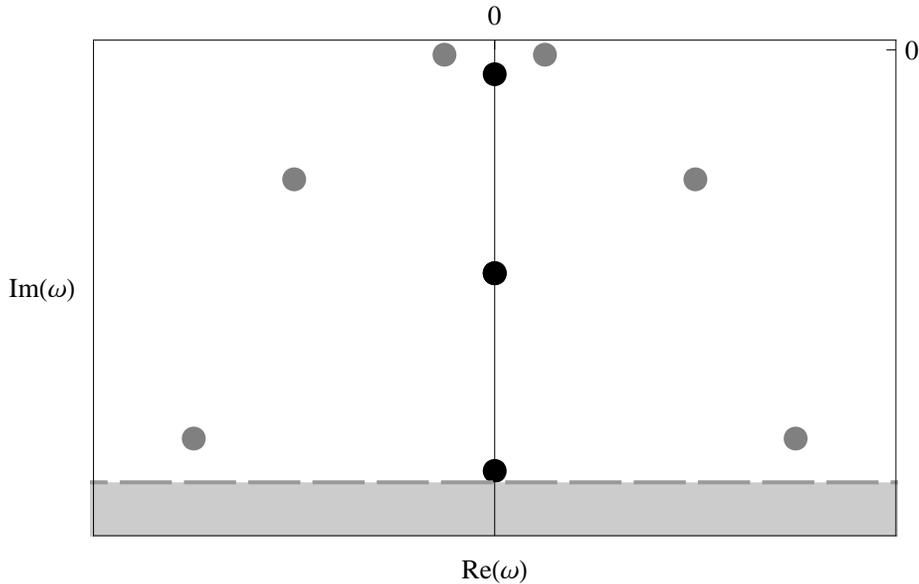} 
\end{tabular}
\caption{\label{fig:polesSchematic} 
Schematic plot of the poles in the coupled system,
i.e. in the broken phase at small finite momentum right below $T_c$. 
These poles are present in each retarded correlation function
for the coupled fields $\eta,\,\sigma,\,A_t,\,A_x$, 
while their residues might vanish for specific fields.
Close to the origin we find the (pseudo)diffusion mode and two hydrodynamic 
second sound modes.
In addition two sets of higher (non-hydrodynamic) quasinormal modes 
are shown. In the unbroken phase these poles originate from the
scalar~(grey dots). We also expect a tower of purely imaginary poles
stemming from the longitudinal vector channel~(black dots).
The grey area indicates where our present numerical methods
break down. In this paper we will be concerned primarily with the hydrodynamic modes
and will touch upon the higher quasinormal modes only briefly.}
\end{figure}

In this section we will apply our determinant method for finding quasinormal
modes of a coupled system of field equations. With this technique we 
follow the model analyzed in the previous section into its broken phase.
Figure \ref{fig:polesSchematic} schematically summarizes our analysis. 
The two formerly separate sets of scalar~(grey dots) and longitudinal vector 
poles~(black dots) present
in the unbroken phase where the scalars and vectors decouple, are now
unified into one inseparable pole structure in the coupled system. 
This is in analogy to a coupled system of two harmonic oscillators in which 
it makes no sense to ask for the eigenfrequencies of the single oscillators.
One could of course try to diagonalize the system of differential equations,
in our case however this looks rather complicated and we prefer to work directly
with the coupled system and with the gauge fields instead of gauge invariant variables.
The lowest modes~(see figure \ref{fig:polesSchematic}) are two 
hydrodynamic second sound modes originating 
from the two lowest scalar quasinormal modes in the unbroken phase.
In addition we will find a non-hydrodynamic pseudodiffusion mode
staying on the imaginary axis in the range of momenta we consider.
This mode can be thought of as the prolongation of the diffusion mode into the 
unbroken phase.

%__________________________________________________________
\subsection{Application of the determinant method}
The equations of motion in the broken phase couple the scalar 
fluctuations~$\eta,\,\sigma$ to the longitudinal vector 
components~$a_t,\, a_x$
\begin{eqnarray}\label{eq:coupledEOMsBroken}
0 &=&  f \eta''
 +\left(f'+\frac{2f}{\rho}\right) \eta'+\left(\frac{\phi^2}{f}+\frac{2}{L^2}
 +\frac{{ \omega}^2}{f}-\frac{{ k}^2}{\rho^2}\right) \eta 
 -\frac{2 i  \omega \phi }{f}\sigma
 -\frac{i\omega \psi}{f}a_t-\frac{i  k \psi}{r^2}a_x  \, , \nonumber \\
 &&\\
0 &=&  f \sigma''
 +\left(f'+\frac{2f}{\rho}\right) \sigma'+\left(\frac{\phi^2}{f}+\frac{2}{L^2}
 +\frac{{ \omega}^2}{f}-\frac{{ k}^2}{\rho^2}\right) \sigma 
 +\frac{2  \phi\psi }{f} a_t+\frac{2 i  \omega \phi }{f}\eta \, , \\
0 &=& f {a_t}''+\frac{2f}{\rho}{a_t}'-\left(\frac{{ k}^2}{\rho^2}+2\psi^2\right){a_t}
-\frac{ \omega k}{\rho^2}a_x-2 i  \omega\psi\,\eta-4\psi\phi\,\sigma  \, , \\
0 &=& f {a_x}'' +f'{a_x}'+\left(\frac{{ \omega}^2}{f}-2\psi^2\right){a_x}
+\frac{ \omega  k}{f}a_t +2 i  k\psi\, \eta\, . 
\end{eqnarray}
This system of four coupled equations is subject to  the constraint 
\begin{equation}\label{eq:constraintBroken}
\frac{ \omega}{f}{a_t}'+\frac{k}{\rho^2}{a_x}'=2i\left(\psi'\,\eta-\psi\,\eta'\right)\, ,
\end{equation}
where the left hand side known from the unbroken phase is amended
by the condensate terms on the right. Note that the real part~$\sigma$ of the 
scalar fluctuation is not involved in the constraint.

The constraint can be interpreted as the Ward identity of current conservation in the
presence of the condensate. We expand the gauge fields near the boundary and note
that $a_0' = \langle n \rangle / \rho^2$ and $a_x' = -\langle j_x \rangle / \rho^2$,
where $j_x$ is the x-component of the current.
Expanding also the r.h.s. and comparing the leading orders in $\rho$ we find
\begin{equation}
\partial_\mu \langle j^\mu \rangle  = 2 \langle O_i \rangle \eta^i_0
\end{equation}
where $\eta^i_0$ is the source for the insertion of the imaginary part of the operator $O_i$, i.e. the Goldstone field
in the dual field theory. This equation is to be understood as the local Ward identity encoding current conservation in the presence of the condensate $\langle O_i \rangle$. It follows for example that the two point function
of the divergence of the current with the operator $O^{(\eta)}_i$ is zero only up to a contact term
$\langle \partial_\mu  j^\mu(x) O^{(\eta)}_i(y) \rangle = \langle O_i\rangle\delta(x-y)$.

The gauge field component~$a_y$ being transverse to the momentum decouples 
from the above system and assumes the form
\begin{eqnarray}\label{eq:transVecEOMBroken}
0&=& f\,{a_y}'' +f'{a_y}'+\left(\frac{{\omega}^2}{f}
-\frac{{ k}^2}{\rho^2}-2\psi^2\right)\,a_y \, .
\end{eqnarray}
Since we do not expect any hydrodynamic modes in the transverse vector channel we
will not study this equation further.

Applying the indicial procedure to the system (\ref{eq:coupledEOMsBroken}) to find 
the exponents for the singular and 
the coefficients for the regular parts of the fields, we obtain the following
behavior at the horizon
\begin{eqnarray}\label{eq:horizonExpansionBroken}
\eta&=& (\rho-1)^\zeta \left(\eta^{(0)}+\eta^{(1)}(\rho-1)
% +\eta^{(2)}(r-1)^2+\eta^{(3)}(r-1)^3
 +\dots\right) \, ,\\
\sigma&=& (\rho-1)^\zeta \left(\sigma^{(0)}+\sigma^{(1)}(\rho-1)
% +\sigma^{(2)}(r-1)^2+\sigma^{(3)}(\rho-1)^3
 +\dots\right) \, ,\\
a_t&=&(\rho-1)^{\zeta+1} \left(a_t^{(0)}+a_t^{(1)}(\rho-1)
% +a_t^{(2)}(r-1)^2+a_t^{(3)}(r-1)^3
 +\dots\right) \, ,\\
a_x&=&(\rho-1)^\zeta \left(a_t^{(0)}+a_t^{(1)}(\rho-1)
% +a_t^{(2)}(r-1)^2+a_t^{(3)}(r-1)^3
 +\dots\right) \, ,
 %,\\
%a_y&=&(r-1)^\zeta \left(a_t^{(0)}+a_t^{(1)}(\rho-1)
% +a_t^{(2)}(r-1)^2+a_t^{(3)}(r-1)^3
% +\dots\right) \, ,
\end{eqnarray}
with the exponent $\zeta=-i \omega/3$ obeying the incoming wave 
boundary condition.

Due to the constraint we can choose only three of the four parameters at the horizon. 
Using the constraint and without loss of generality we can eliminate the time 
component~$a^0_t$ and 
parametrize the solutions by $(\eta^{(0)},\sigma^{(0)},a_x^{(0)})$. We choose
three linearly independent combinations~$I, II, III$.
A fourth solution can be found from the gauge transformations 
\begin{equation}\label{eq:soltrivialbroken}
\eta^{IV}=i \lambda \psi \, ,\quad
\sigma^{IV}= 0\, ,\quad
a_t^{IV}= \lambda\omega\,,\quad 
a_x^{IV}=-\lambda k\, .
\end{equation}
with $\lambda$ being an arbitrary constant with respect to $\rho$.
It is not an algebraic solution to the equations of motion since $\eta$ has non trivial
dependence on the bulk variable $\rho$. The gauge solution solves the equations 
(\ref{eq:coupledEOMsBroken})
not exactly but only up to terms proportional to the background equations 
(\ref{eq:background}).

Our goal is to find the poles in the retarded correlation functions 
of the four fields appearing in the coupled system of equations
of motion~\eqref{eq:coupledEOMsBroken}. A convenient way of imposing
the appropriate boundary conditions is given by redefining the scalar
fields as
\begin{equation}
\tilde\eta(\rho) = \rho\eta(\rho) ~~~~,~~~~ \tilde\sigma(\rho) = \rho\sigma(\rho)\,.
\end{equation}

Then the most general solution for each field~$\varphi_i\in\{\tilde\eta,\tilde\sigma,a_t,a_x\}$ 
including gauge degrees of freedom can be written 
\begin{eqnarray}\label{eq:generala}
\varphi_i=\alpha_1{\varphi_i}^I+\alpha_2{\varphi_i}^{II}
 +\alpha_3{\varphi_i}^{III}+\alpha_4{\varphi_i}^{IV} \, .
\end{eqnarray}

In the theory with the dimension two operator the sources for the various
gauge invariant operators are given
by $\varphi_i(\Lambda)$.
We are interested in the quasinormal modes of the system (\ref{eq:coupledEOMsBroken})
and as we have argued in the previous section these are the special values of the frequency
where the determinant spanned by the values $\varphi^{I,II,III,IV}_i$ vanishes.
Expanding this determinant we get
\begin{eqnarray} \label{eq:detBroken}
0 &=& \frac{1}{\lambda}\det \left(
\begin{array}{cccc}
{\varphi_\eta}^I&{\varphi_\eta}^{II}&{\varphi_\eta}^{III}&{\varphi_\eta}^{IV} \\
{\varphi_\sigma}^I&{\varphi_\sigma}^{II}&{\varphi_\sigma}^{III}&{\varphi_\sigma}^{IV} \\
{\varphi}_t^I&{\varphi}_t^{II}&{\varphi}_t^{III}&{\varphi}_t^{IV} \\
{\varphi}_x^I&{\varphi}_x^{II}&{\varphi}_x^{III}&{\varphi}_x^{IV} \\
\end{array}
\right) \, \\
&=& i \varphi^{IV}_\eta \det\left(
\begin{array}{ccc}
{\varphi_\sigma}^I&{\varphi_\sigma}^{II}&{\varphi_\sigma}^{III} \\
{\varphi}_t^I&{\varphi}_t^{II}&{\varphi}_t^{III} \\
{\varphi}_x^I&{\varphi}_x^{II}&{\varphi}_x^{III}\\
\end{array}
\right)
+\omega \det \left(
\begin{array}{ccc}
{\varphi_\eta}^I&{\varphi_\eta}^{II}&{\varphi_\eta}^{III}\\
{\varphi_\sigma}^I&{\varphi_\sigma}^{II}&{\varphi_\sigma}^{III} \\
{\varphi}_x^I&{\varphi}_x^{II}&{\varphi}_x^{III}\\
\end{array}
\right)
+k \det \left(
\begin{array}{ccc}
{\varphi_\eta}^I&{\varphi_\eta}^{II}&{\varphi_\eta}^{III}\\
{\varphi_\sigma}^I&{\varphi_\sigma}^{II}&{\varphi_\sigma}^{III} \\
{\varphi}_t^I&{\varphi}_t^{II}&{\varphi}_t^{III}\\
\end{array}
\right) \, , \nonumber 
\end{eqnarray}
which needs to be evaluated at the cutoff ${\rho=\Lambda}$. 
The first term in \eqref{eq:detBroken} vanishes at the cutoff since 
$\varphi^{IV}_4=\Lambda \psi=0$ is just
the condition that the operator $O_2$ is not sourced by the background. 

We first find three linearly independent numerical solutions and then 
solve condition~\eqref{eq:detBroken} numerically. Explicit checks confirm that the 
choice of a solution basis $\varphi^{I,II,III,IV}$ is completely arbitrary
and does not change the results. Note also that in our present case
all the remaining determinants can not be factorized. But if the momentum
is set to zero, the only remaining term is the one with $\omega$ and the 
determinant factorizes into a scalar part and a vector part
since the system of equations decouples. 

In the theory with the dimension one operator the sources are given by 
$-\Lambda^2\tilde\eta$ and $-\Lambda^2\tilde\sigma$ for the scalar fields.
Let us call $\varphi_1=-\rho^2 \tilde\eta'$ and $\varphi_2=-\rho^2 \tilde\sigma'$
in this case. The determinant has therefore the same form and again the first term
vanishes due to the absence of sources for $O_1$ in the background solution.
The quasinormal modes can again be found by integrating three arbitrary solutions
with infalling boundary conditions from the horizon to the cutoff and finding
numerically the zeroes of the determinant (\ref{eq:detBroken}).

%__________________________________________________________
\subsection{Hydrodynamic and Goldstone modes}

\begin{figure}[!htbp]
\psfrag{Rew}{$\text{Re}\,\omega$}
\psfrag{Imw}{$\text{Im}\,\omega$}
\psfrag{k}{$ k$}
\centering
\begin{tabular}{c}
\includegraphics[width=0.8\textwidth]{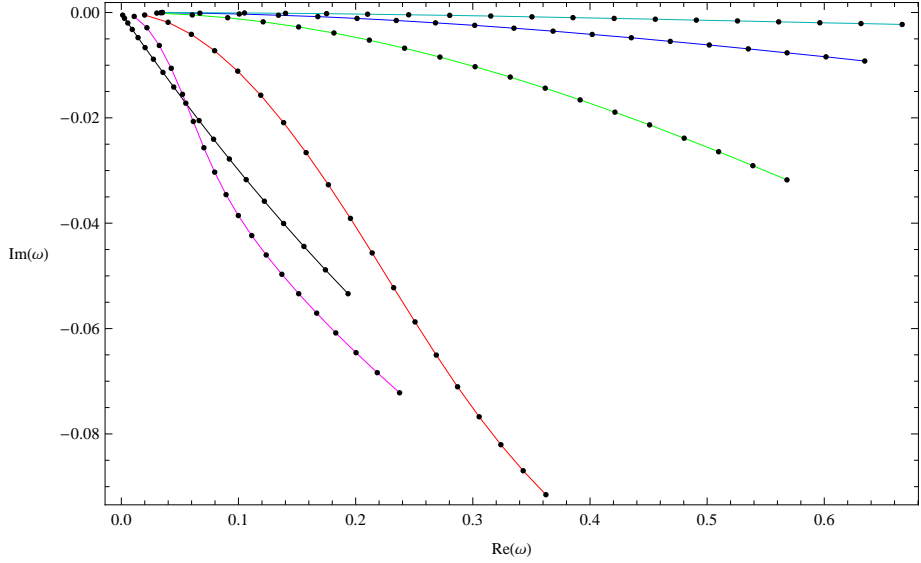} 
\end{tabular}
\caption{\label{fig:soundQNM} Movement of the positive frequency 
sound pole away from $\omega=0$ with increasing
spatial momentum. Distinct curves correspond to temperatures below
the phase transition $T/T_c=0.999$~(black), $0.97$~(pink), $0.91$~(red), 
$0.71$~(green), $0.52$~(blue), $0.26$~(light blue). Dots on one curve are 
separated by $\Delta  k=0.05$. All curves start at $ k=0.05$ and
end at $ k=1.00$.
 }
\end{figure}

\paragraph{Sound mode}
The scalar modes originally destabilizing the unbroken phase turn into 
Goldstone modes at $T_c$ instead of becoming tachyonic. Below $T_c$
they evolve into the two second sound modes.
Figure~\ref{fig:soundQNM} shows their movement when momentum
is changed at different temperatures. 
Note that we focus on the positive real frequency axis because of 
the mirror symmetry sketched in figure \ref{fig:polesSchematic}.
From the dispersion relation at small frequencies and long wavelengths we extract the 
speed of second sound~$v_s$ and the second sound 
attenuation~$\Gamma_s$ using the hydrodynamic equation
\begin{equation}\label{eq:disprelvs}
\omega= {v_s} k-i\Gamma_s k^2\, .
\end{equation}
\\

\begin{figure}[!htbp]
\centering
\begin{tabular}{cc}
\includegraphics[width=7cm]{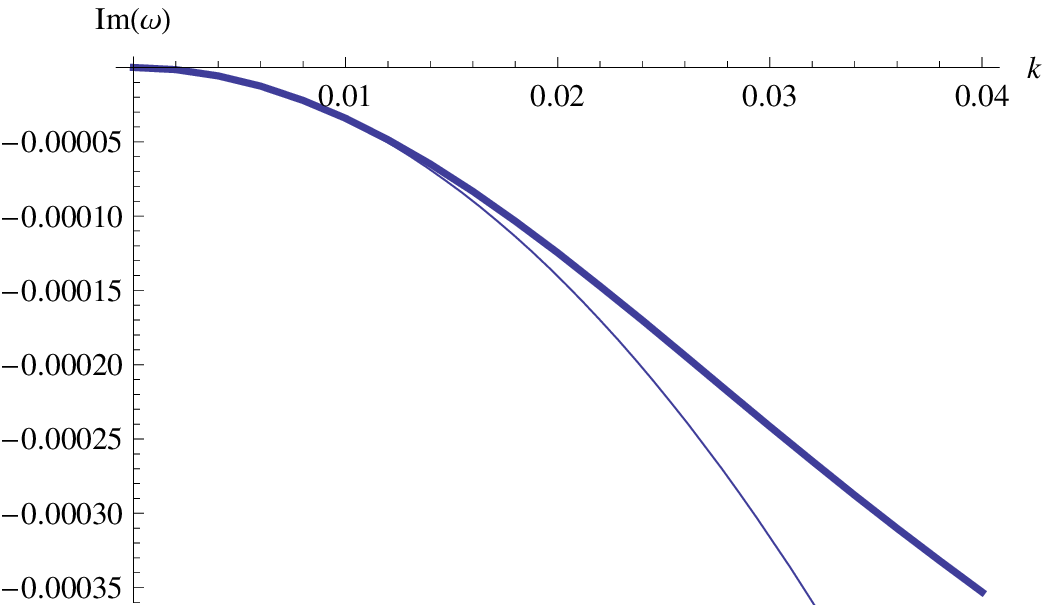}  & \includegraphics[width=7cm]{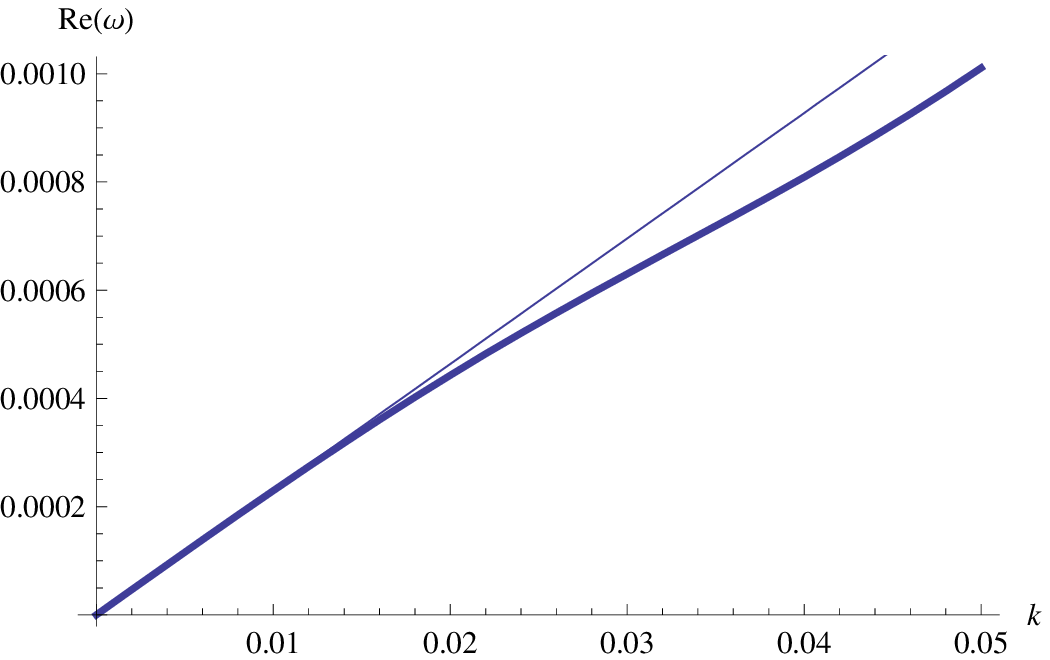}  
\end{tabular}
\caption{\label{fig:disprelfits} Fits of the real and imaginary part of the hydrodynamic modes in the broken phase
to the lowest order approximation $\omega=v_s k -i \Gamma_s k^2$. The left figure shows the real part and the right one
the imaginary part. The thick lines are the numerical results and the thin lines are the linear and quadratic fits. 
The fit is done for a temperature just below the 
critical one where the range of the approximation is rather small.}
\end{figure}

It turns out that the hydrodynamic regime, i.e. that range of momenta in which the
dispersion relation is well approximated by (\ref{eq:disprelvs}), is very narrow
for temperatures just below the critical one since the speed of sound vanishes
at $T_c$. Fits to the hydrodynamic form at a high temperature $T\approx0.9999 T_c$
are plotted in figure \ref{fig:disprelfits} for the $O_2$-theory, the results for
the $O_1$ theory are qualitatively similar.

\begin{figure}[!htbp]
\centering
\begin{tabular}{cc}
\includegraphics[width=7cm]{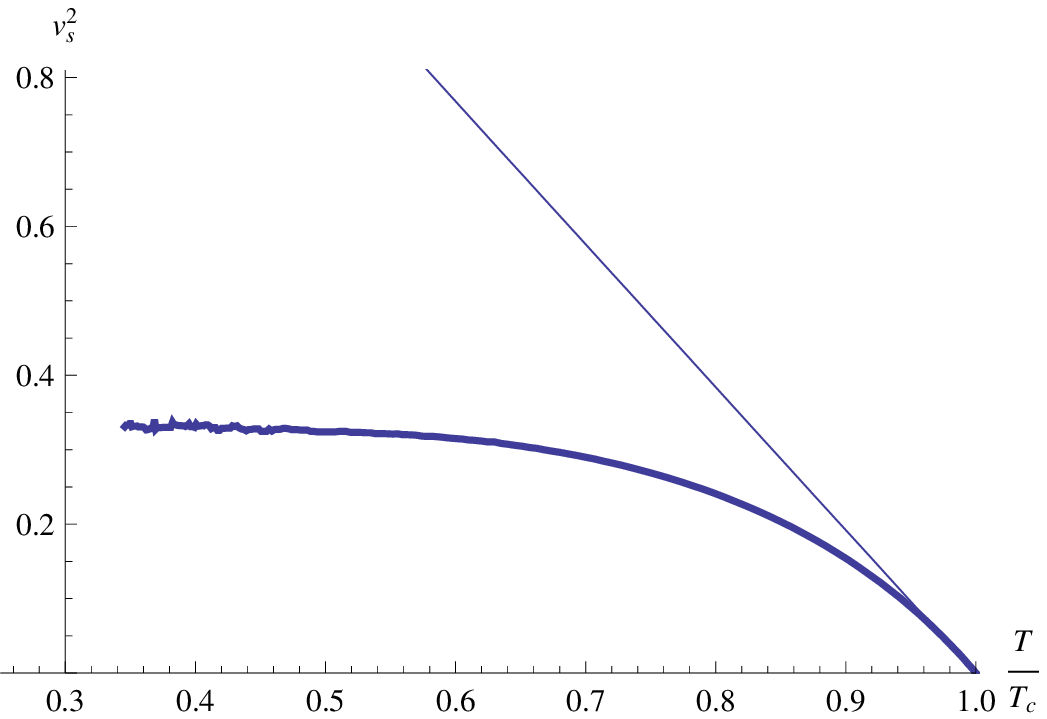} & \includegraphics[width=7cm]{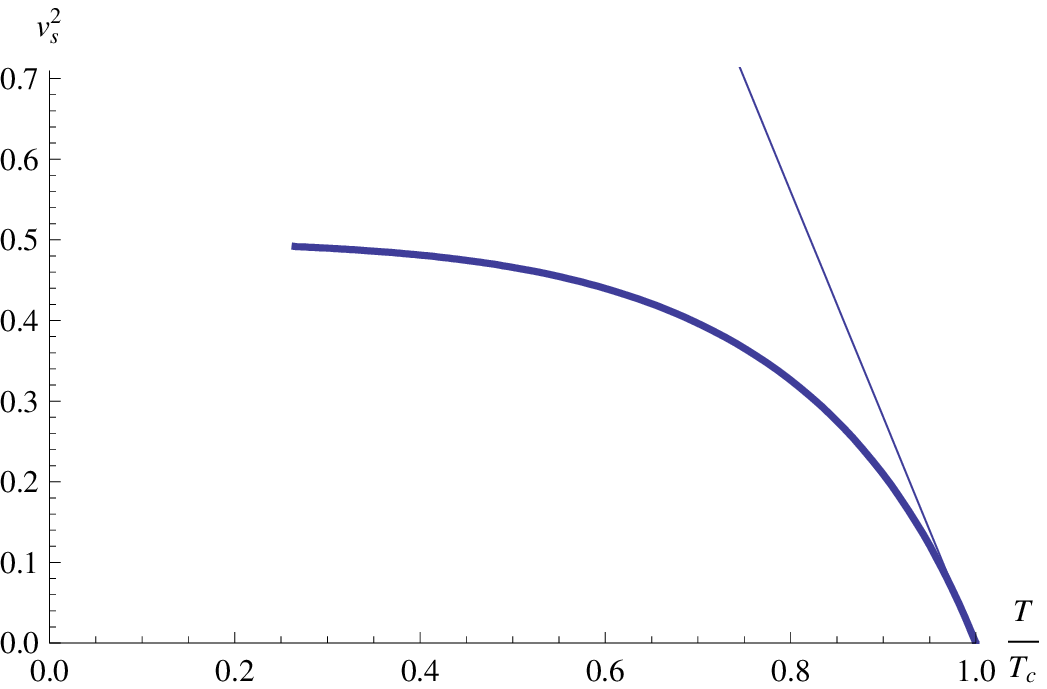} 
\end{tabular}
\caption{\label{fig:speedofsound}The plots show the squares of the speed of sound as extracted from the location of the lowest
quasinormal mode in the broken phase. The left figure is for the $O_1$ theory and the right one for the $O_2$ theory. We also
indicate the linear behavior close to $T_c$.  As can be seen the numerics for the $O_1$ theory becomes somewhat unstable for low
temperatures.}
\end{figure}

The speed of second sound is shown in figure~\ref{fig:speedofsound}.
We have a good numerical agreement with the thermodynamic value of the second 
sound velocity given in~\cite{Herzog:2008he}. This nicely confirms 
validity of our method. In particular, within our numerical 
precision we find that the value of the square of the speed of sound tends to $v_s^2\approx 1/3$ 
in the $O_1$ theory and to $v_s^2\approx 1/2$ in the $O_2$ theory\footnote{In \cite{Yarom:2009uq} it was argued
that conformal symmetry implies $v_s^2=1/2$ at zero temperature. Due to the divergence in the order parameter for
the $O_1$ theory conformal symmetry could be broken and allow thus for a different value.} 
. However, the numerics becomes
rather unstable for low temperatures, especially for the $O_1$ theory.
Near but below the critical temperature we find 
\begin{eqnarray}
v_s^2 &\approx & 1.9 \left(1-\frac{T}{T_c}\right) ~~~~\mathrm{O_1-Theory}\,,\\
v_s^2 &\approx & 2.8 \left(1-\frac{T}{T_c}\right) ~~~~\mathrm{O_2-Theory}\,.
\end{eqnarray}

\begin{figure}[!htbp]
\centering
\begin{tabular}{cc}
\includegraphics[width=7cm]{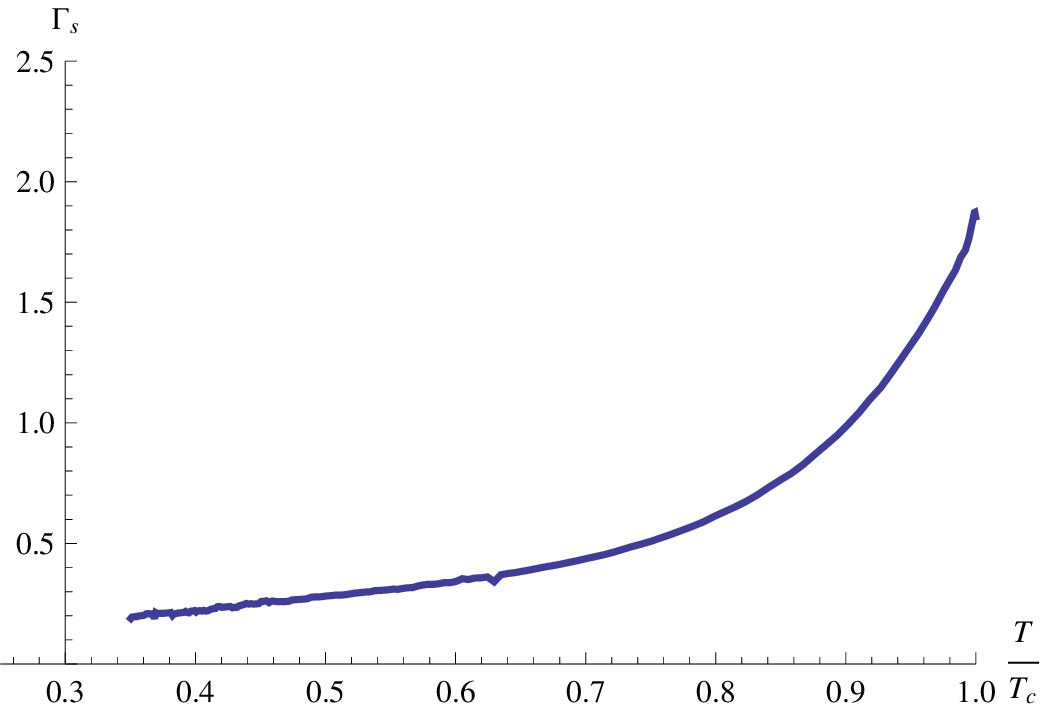} & \includegraphics[width=7cm]{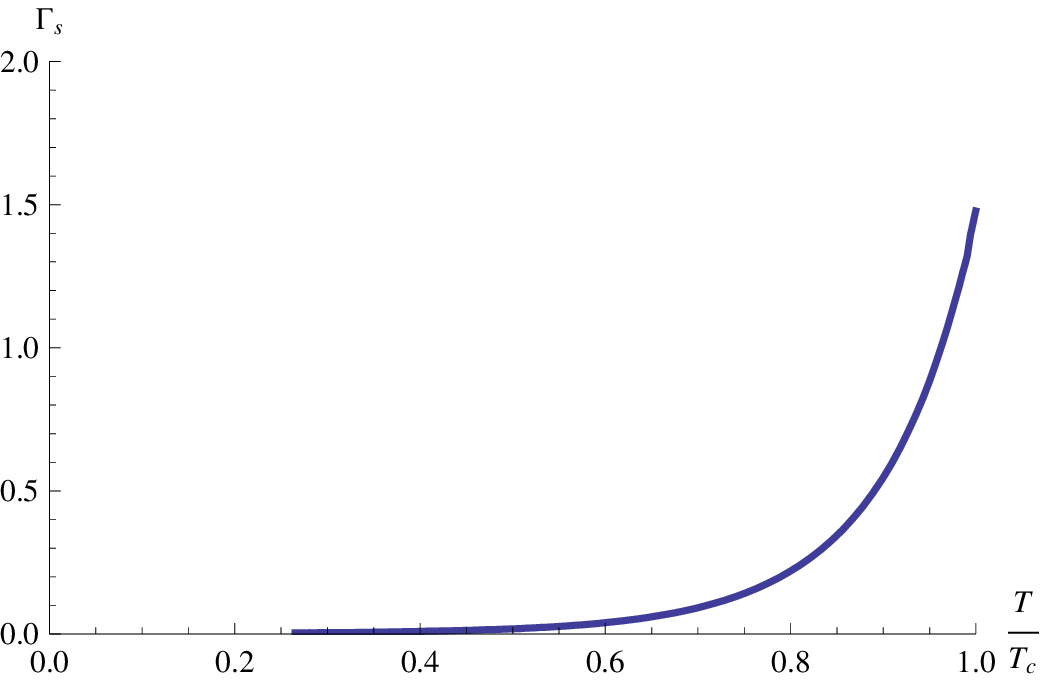} 
\end{tabular}
\caption{\label{fig:soundattenuation}The plots show the attenuation constants of second sound as extracted from the location of the lowest
quasinormal mode in the broken phase. The left figure is for the $O_1$ theory and the right one for the $O_2$ theory. 
Again it can be noticed that the $O_1$ theory is numerically more challenging at low temperatures.}
\end{figure}

Moreover, as a benefit of our effort considering the fluctuations, we are also 
able to extract non-thermodynamic quantities in this channel. Specifically
we examine the attenuation of the second sound mode as shown 
in figure \ref{fig:soundattenuation}. The curve  shows
how attenuation smoothly asymptotes to zero as the superfluid becomes more
and more ideal at low temperatures. This effect is however much stronger in
the $O_2$ theory. Near $T_c$ the attenuation is growing. Within our numerical
precision it seems however that the attenuation constant is taking a finite value
at the critical temperature. Numerically we find 
\begin{eqnarray}
\Gamma_s &=& 1.87 T_c ~~~\mathrm{at}~~~ T=0.9991 T_c ~~~~\mathrm{O_1-Theory}\,,\\
\Gamma_s &=& 1.48 T_c ~~~\mathrm{at} ~~~T=0.9998 T_c ~~~~\mathrm{O_2-Theory}\,.
\end{eqnarray}
A similar behavior has been observed in~\cite{Buchel:2008uu}, where 
the attenuation of the normal sound mode asymptotes to a finite value near 
a phase transition.

\paragraph{(Pseudo) diffusion mode}
The vector diffusion mode from the unbroken phase turns into a
(pseudo) diffusion mode below $T_c$\footnote{An analytical result obtained for second sound in a non-abelian
model~\cite{Herzog:2009ci} also shows the appearance of a pseudo diffusion mode
with a gap that vanishes as the condensate goes to zero.}. 
For not too low temperatures and 
not too large momenta the dispersion relation for this mode is well approximated by  
\begin{equation}\label{eq:pseudoDiffusion}
\omega=-iDk^2-i\gamma(T) \,,
\end{equation}
with a gap $\gamma\in\mathbb{R}$ in imaginary frequency direction.
Thus the pole is shifted from its unbroken phase position such that it does 
not approach zero at vanishing momentum any more,  i.e. it is not anymore
a hydrodynamic mode.

\begin{figure}[!htbp]
\centering
\begin{tabular}{cc}
\includegraphics[width=7cm]{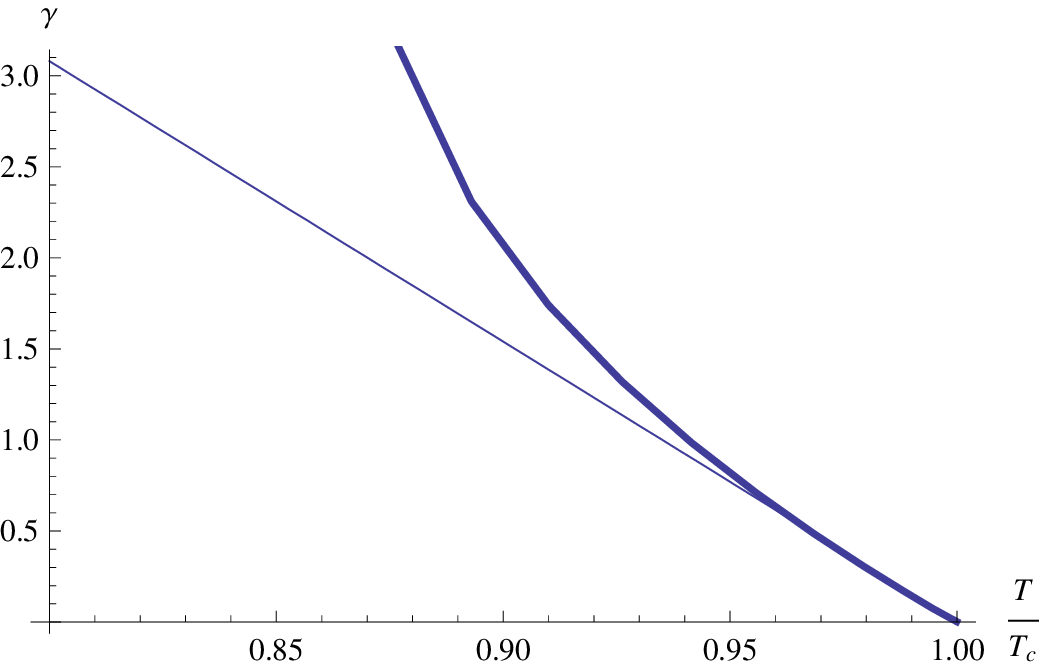} & \includegraphics[width=7cm]{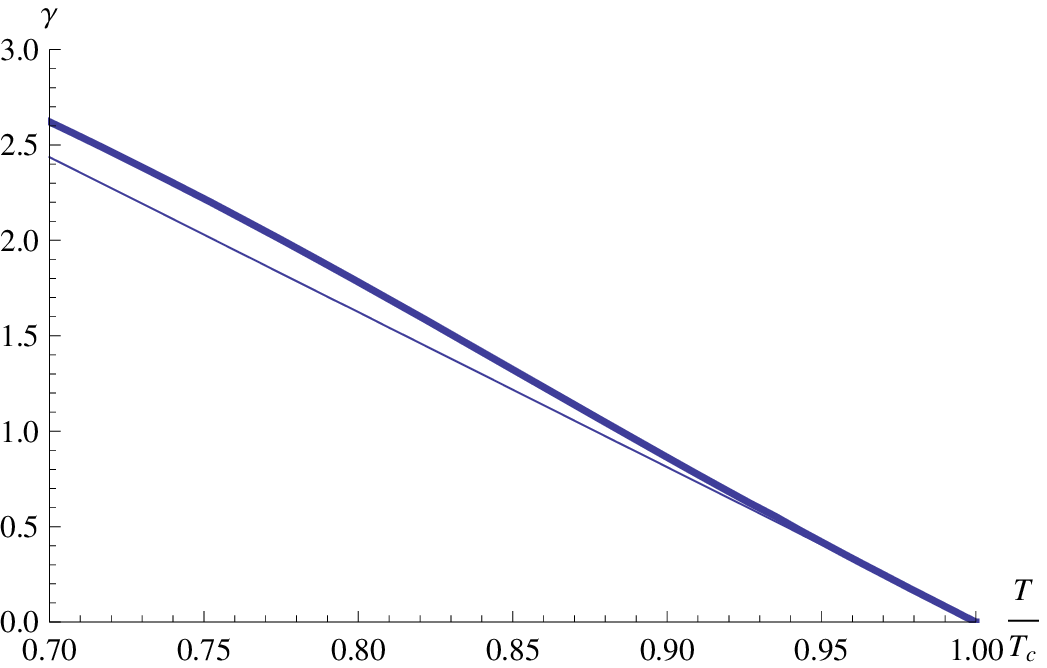} 
\end{tabular}
\caption{\label{fig:diffusiongap}The plots show the gap of the pseudo diffusion mode as a function of reduced temperature.
On the left the $O_1$ theory and on the right the $O_2$ theory. }
\end{figure}

In figure \ref{fig:diffusiongap} we have plotted the gap $\gamma$
as a function of the reduced temperature and we can see that it vanishes
linearly near $T_c$. 
\begin{eqnarray}\label{eq:critScalDiffGap}
\gamma &\approx & 15.4 T_c\left(1-\frac{T}{T_c}\right) ~~~~\mathrm{O_1-Theory}\, , \\
\gamma &\approx &  8.1 T_c\left(1-\frac{T}{T_c}\right) ~~~~\mathrm{O_2-Theory}\,. 
\end{eqnarray}

\begin{figure}[!htbp]
\centering
\begin{tabular}{cc}
\includegraphics[width=7cm]{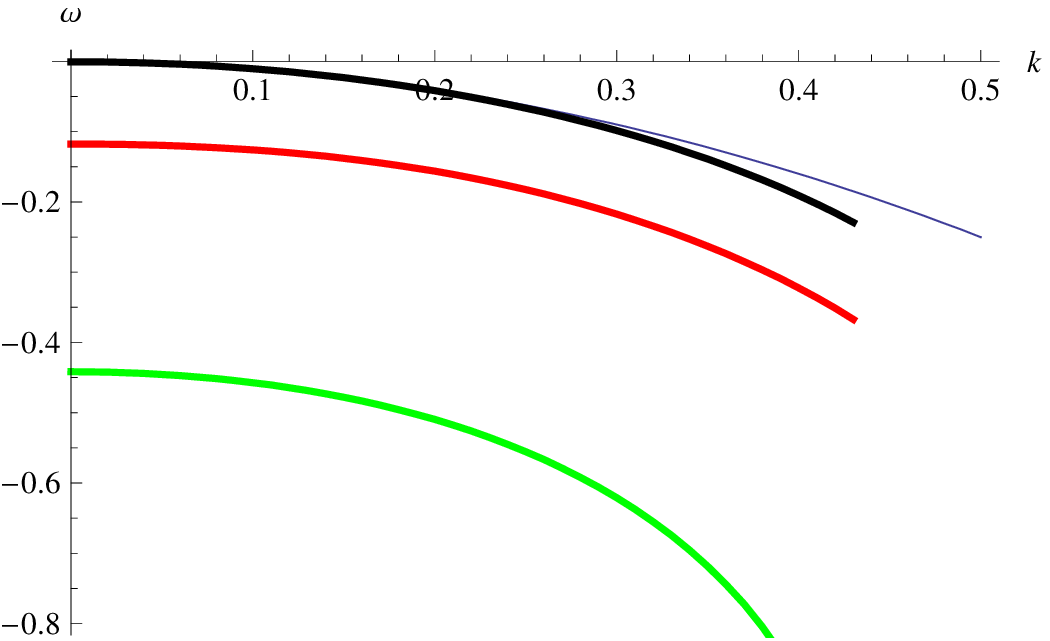} & \includegraphics[width=7cm]{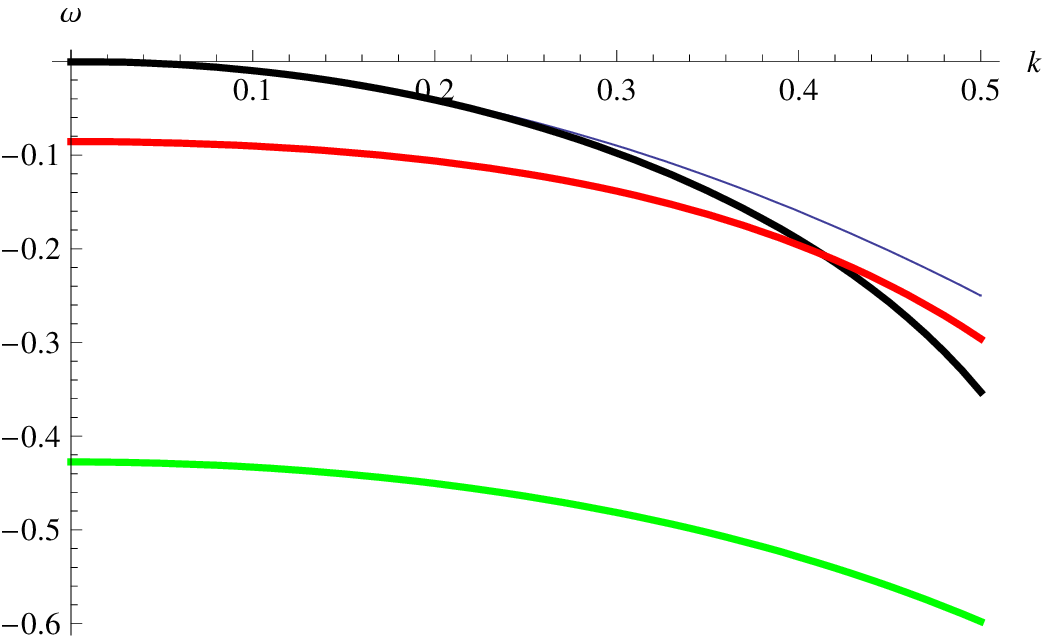} 
\end{tabular}
\caption{\label{fig:diffusionQNM}The plots show the dispersion relations for the
pseudo diffusion mode at different temperatures and the diffusion dispersion relation with $D=3/(4\pi T)$ (note that
even in the unbroken phase the latter approximates the diffusive quasinormal mode only for small momenta). 
On the left we have the $O_1$ theory at
temperatures $T=0.999 T_c$ (black), $T=0.97Tc$ (red) and $T=0.91T_c$ (green). On the right the same for the $O_2$ theory
at temperatures $T=0.999 T_c$ (black), $T=0.97T_c$ (red) and $T=0.87T_c$ (green).}
\end{figure}

Figure~\ref{fig:diffusionQNM} shows the dispersion relation for the 
diffusion pole at different temperatures. The offset at $ k=0$ is the
gap size~$\gamma$ depending linearly on $T$ only near $T_c$.
This implies that the relation~\eqref{eq:pseudoDiffusion} asymptotes
to the ordinary diffusion equation near the critical temperature.
As expected the highest temperature curve $T=0.999T_c$~(black) matches
the hydrodynamic approximation (thin line) very well at small momenta.
That agreement becomes worse around $ k\sim0.25$. Also as the 
condensate grows below $T_c$ the behavior of this (pseudo) diffusion 
mode becomes less hydrodynamic.
\\

%__________________________________________________________
\subsection{Higher quasinormal modes}

In addition to the hydrodynamic sound modes and the pseudo diffusion mode 
there are higher quasinormal modes. They are not the main focus of 
this paper and we have not studied them in detail. 
We have however traced the prolongation of the second and third quasinormal
modes in the scalar sector from the unbroken phase into the broken phase.

\begin{figure}[!htbp]
\centering
\begin{tabular}{c}
\includegraphics[width=0.8\textwidth]{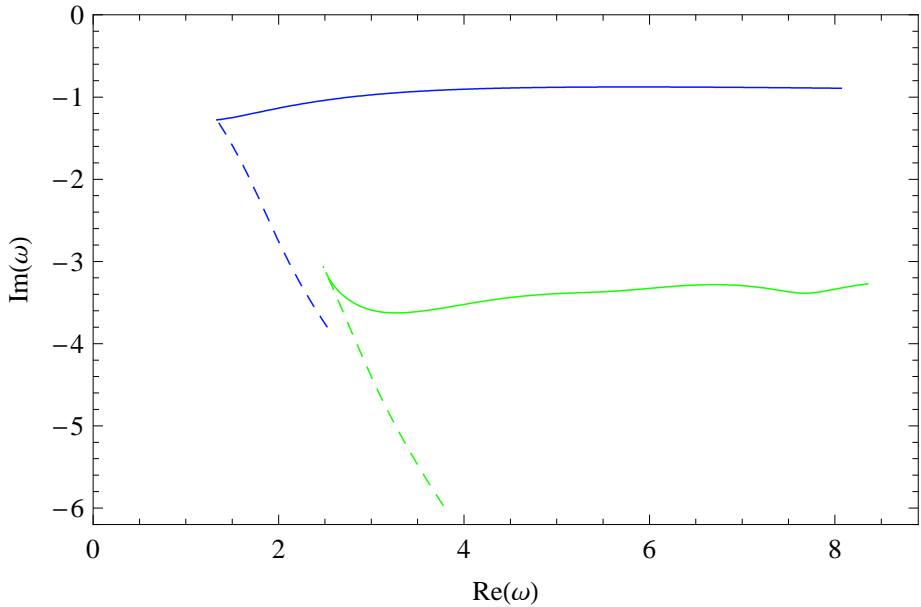} 
\end{tabular}
\caption{\label{fig:hiQNMs} Movement of the higher poles
in the complex frequency plane at vanishing momentum $ k=0$. 
Scalar modes 2 (blue) and 3 (green) in the
unbroken phase (dashed) evolve continuously into the higher
poles of the broken phase (solid). The right end points are evaluated for $T/T_c=0.25$.}
\end{figure}

The former scalar modes evolve continuously into higher modes of
the coupled system through the phase transition as seen from the
two kinked dashed (unbroken phase) and solid (broken phase) lines 
in figure~\ref{fig:hiQNMs}. The kink 
indicates that the poles move continuously but change direction at the
critical temperature. We show only the plot for the theory with the dimension
two operator, similar results hold for the $O_1$ theory. At the critical
temperature the locations of the quasinormal frequency calculated with the
Frobenius method in the unbroken phase and by the method of finding the
zeroes of the determinant spanned by the solutions match with impressively
high precision. We might take this as a highly non trivial test of the 
accuracy of the numerical integration method.

We expect that all quasinormal frequencies are shifted {\it continuously} in
the complex frequency plane across the phase transition. 
This means that there are no jumps in any of the dispersion relations. 
There is simply an infinite set of poles corresponding to the degrees of freedom
of the system which are continuously shifted when parameters are changed.

Our numerical invesigations also reveal poles developing
an increasing real part with decreasing temperature which most likely
originate from the longitudinal vector modes in the unbroken phase.
Note that we have not shown these modes for simplicity.

The two higher modes shown in figure~\ref{fig:hiQNMs} move 
parallel to the real axis and to each other with decreasing 
temperature.~\footnote{Note that this behavior is very different from
the behavior usually found in five-dimensional holographic setups when temperature 
is decreased~\cite{Myers:2007we,Erdmenger:2007ja,Paredes:2008nf}.} 
At low temperatures their real parts are almost the same. This is also true
for the longitudinal vectors as far as we can tell within our numerical
uncertainties. We suspect that this {\it alignment of higher poles} has to do with 
the appearence of the conductivity gap showing up at low temperatures~\cite{Hartnoll:2008vx}.

%%%%%%%%%%%%%%%%%%%%%%%%%%%%%%%%%%%%%%
%%%%%%%%%%%% DISCUSSION %%%%%%%%%%%%
%%%%%%%%%%%%%%%%%%%%%%%%%%%%%%%%%%%%%%
\section{\label{sec:discussion}Summary and Discussion}

We have studied the hydrodynamics of a holographic model of a superfluid.
The model is an Abelian gauge field model with a charged scalar field in a
four dimensional AdS black hole background.

As is well known by now the holographic dictionary allows to interpret the quasinormal
frequencies as the poles of the retarded Green functions of the dual field theory.
 By calculating the low lying quasinormal frequencies numerically in the broken and
 unbroken phases we were able to identify the hydrodynamics. We found that at
 high temperatures, $T>T_c$ there is only one hydrodynamic mode representing the
 diffusion of the conserved $U(1)$ charge. As one lowers the temperature reaching the
 critical temperature $T_c$ two of the quasinormal frequencies of the charged scalar field
 approach the origin of the complex frequency plane. Precisely at the critical temperature at
 the onset of the phase transition these modes become massless, giving rise to new hydrodynamic
 variables. We also have calculated the residue of these modes and found that it stays finite at
 $T=T_c$ resulting in a divergence of the order parameter susceptibility, as expected.
 
 Below the critical temperature these modes stay massless and show a dispersion relation with
 a linear real part and a quadratic imaginary one, allowing an interpretation as the
 modes of second sound in the superfluid. On the other hand the diffusion mode  starts to develop
 a gap and stops to be hydrodynamic. The counting is therefore one hydrodynamic mode at
 high temperatures, three at the critical temperature and two at low temperatures.

 In the low temperature phase we were able to calculate the speed of sound as well as its attenuation
 constant as a function of temperature. Also the gap in the pseudo diffusion mode has been determined.
 
 We have also been able to follow some of the higher quasinormal modes through the phase transition and
 found that they evolve continuously albeit non-smoothly with temperature in the complex frequency plane, showing
 a sharp kink at the critical temperature.
 
 On a technical side we have developed a method to determine the quasinormal frequencies and the
 holographic Green functions for systems of coupled differential equations and without using gauge invariant
 variables. The quasinormal frequencies correspond simply to the zeroes of the determinant spanned
 by a maximal set of linearly independent solutions. We have furthermore seen that the poles of the Green
 functions stemming from bulk gauge fields are gauge invariant as expected. 
 
 There are now several interesting questions that should be investigated in the future. The most obvious one
 is to apply the methods developed here to the model where the backreaction of the matter and gauge field
 onto the metric are properly taken into account. Due to the presence of the metric fluctuations the hydrodynamics
 of such a model is certainly much richer. In addition to the diffusion and second sound modes found here one
 expects also shear and sound modes stemming from bulk metric fluctuations. The corresponding system
 of differential equations promises to be rather involved. However there should be no principal obstacle to apply
 our methods also in these cases. Such an investigation is currently underway \cite{progress}. 
  
 Another interesting direction of research should be to reinterpret the results obtained here and analogous ones
 for related models with different scalar mass and living in different dimensions from the point of view of dynamical
 critical phenomena \cite{Hohenberg:1977}. 
 We have seen already that the speed of sound scales with exponent one half, whereas the gap in the pseudo diffusion mode scales with exponent one. The situation for the sound attenuation is unfortunately less clear. As far as our numerics
 indicates the sound attenuation reaches a finite value at $T=T_c$. An extensive study of related models possibly
 with enhanced numerical efforts might give new insights here. 
 
 Of course an effort should also be undertaken to extend the results at hand to models of $p$-wave superconductors \cite{Gubser:2008wv}.
 For infinitesimal condensates and analytic study of the hydrodynamics has already been done in \cite{Herzog:2009ci}. It might
 be of interest to supplement these analytical result with a numeric study that allows to go further away from the
 phase transition point deep into the broken phases. Another very interesting class of holographic $p$-wave superconductors
 are the ones realized on $D7$ brane embeddings \cite{Ammon:2008fc,Basu:2008bh,Ammon:2009fe}. Due to the presence of fundamental matter these should be especially interesting to study.  
 
 We hope to come back to all or at least some of these and other questions in future publications.

%__________________________________________________________

%%%%%%%%%% Acknowledgements %%%%%%%%%%
\begin{acknowledgments}
We thank S.~Hartnoll and S.~Sachdev for discussions.\\
This work has been supported partially by the Plan Nacional de Altas Energ\'ias FPA-2006-05485, FPA-2006-05423, EC Commission under grant MRTN-CT-2004-005104 and Comunidad de Madrid HEPHACOS P-ESP-00346. I.~A. is supported by grant BES-2007-16830.
\end{acknowledgments}

%%%%%%%%%%%%%%%%%%%%%%%%%%%%%%%%%%%%%%
%%%%%%%%%% THE BIBLIOGRAPHY %%%%%%%%%%
%%%%%%%%%%%%%%%%%%%%%%%%%%%%%%%%%%%%%%
\bibliographystyle{unsrt}
\bibliography{HydroSupercond}
\end{document}